\documentclass[11pt,a4paper]{article}
\usepackage[utf8]{inputenc}
\usepackage{amsmath,amssymb,amsthm}
\usepackage{geometry}
\usepackage{microtype}
\usepackage{booktabs}
\usepackage{array}
\usepackage{graphicx}
\usepackage{tikz}
\usetikzlibrary{decorations.pathmorphing,decorations.pathreplacing,arrows.meta,positioning,calc}
\usepackage{hyperref}

\newcommand{\incfig}[2][0.92\textwidth]{\IfFileExists{#2}{\includegraphics[width=#1]{#2}}{\fbox{\parbox[c][4cm][c]{#1}{\centering\ttfamily Figure file \detokenize{#2} not found}}}}

\newtheorem{theorem}{Theorem}
\newtheorem{proposition}{Proposition}
\newtheorem{lemma}{Lemma}
\newtheorem{corollary}{Corollary}
\theoremstyle{definition}
\newtheorem{definition}{Definition}
\theoremstyle{remark}
\newtheorem{remark}{Remark}

\title{\textbf{Why the Swing Parameters Are Sloppy}}

\author{Sunny Kumar\thanks{Postdoc, ECE Dept, Santa Clara University, skumar13@sc.edu}, \ \ Madhavi~Parimi\thanks{Research Scholar, E-MC$^2$ Lab, VJTI, Mumbai, India}, \ \
Shashank Varma\thanks{Lecturer, EED, VJTI}, \ \
Sushama Wagh\thanks{Faculty--EED \& P.I.\ (E-MC$^2$), VJTI},\\ \ \ Aleksandar M. Stankovi\'c \thanks{Professor, ECE Dept, Santa Clara University, astankovic@scu.edu},\ \
Navdeep Singh\thanks{IGI Research Chair Professor, E-MC$^2$ Lab, VJTI},}

\date{}

\begin{document}

\maketitle

\begin{abstract}
When a multi-parameter model is fitted to measured data, only a few parameter combinations are often identifiable, as Fisher Information Matrix eigenvalues span several orders of magnitude. Sensitivity-based tools, including the Fisher Information Matrix and MBAM, reveal this structure analytically, while manifold learning can recover it from sampled trajectories. For an infinite-bus synchronous generator, both approaches show that inertia \(H\) and damping \(D\) are among the least identifiable parameters on a late post-disturbance window \(t\in[3,5]\) s, where six of the eleven parameters suffice to describe the response.
We explain this result from the geometry of the dynamics rather than from that of the parameter-to-data map. The rotor swing is a fast fibre that has collapsed onto a two-dimensional normally attracting invariant manifold (NAIM) of the transient fluxes before the window opens. The zeroth-order NAIM is exactly independent of $H$, $D$, $T_{d0}''$ and $T_{q0}''$, four of the five MBAM limits. After the collapse a fast parameter reaches the record only through a two-dimensional memory channel, which the identifiable parameters mimic to better than $99.6\%$, and through a tilt of the manifold suppressed by the time-scale ratio. We prove, under stated smoothness and rank conditions, that each of the four limits is a boundary of the model manifold at a distance equal, to first order in the parameter, to its current profiled speed. This yields the boundary distances and the reduced model without integrating a geodesic, and the ranking of the distances matches the MBAM sequence; geodesics we computed agree with the predicted lengths to within $5\%$. The same analysis predicts how identifiability changes with the experiment: neither a larger initial speed deviation nor a load change restores $H$ or $D$ in the late window, whereas the early window $[0, 2]\text{ s}$ makes $H$ the best-determined parameter. Evaluated in a physical metric fixed by synchrophasor accuracy limits, the late window identifies five parameters rather than six and the early window nine. The spiked-covariance structure of the linearised trajectory map explains why the principal components of whitened, sampled trajectories reproduce the prior-scaled Fisher spectrum, and regression on those components attains the Fisher-predicted recovery error in 113 of 114 parameter--scenario pairs.
\end{abstract}


\section{Introduction}

\subsection{The paradox}
When a model of a synchronous generator is fitted to a measured trajectory, one hopes that the data pin down every parameter. In practice some parameters can be moved a long way with almost no change in the fit. They are called sloppy. The surprise in the infinite-bus generator is which parameters turn out to be the sloppiest. The inertia constant $H$ and the damping $D$ govern the electromechanical swing, the most visible and most studied motion of the machine. Yet on a record taken a few seconds after a disturbance they are the two least identifiable of the model's eleven parameters. Their Fisher-information eigenvalues lie nine and eleven orders of magnitude below the stiffest one.

A mechanical picture explains this, if it is set up carefully. Linearised about its operating point, the swing equation is the equation of a weight on a spring, with a dashpot alongside:
\begin{equation}
\frac{H}{\omega_b}\ddot{\delta} + \frac{D}{\omega_b}\dot{\delta} + K_s(\delta - \delta^*) \approx 0, \qquad K_s = \frac{\partial P_g}{\partial \delta}.
\label{eq:swing_mech}
\end{equation}
The rotor angle $\delta$ is the position of the weight, $H/\omega_b$ is its mass, $D/\omega_b$ is a drag against the fixed synchronous frame (it acts on the deviation of the speed from synchronous), and the synchronising power coefficient $K_s$ is the stiffness of an electrical spring. The spring is not anchored to the ground. Its anchor is the angle $\delta^*$ at which electrical and mechanical power balance, $P_g(\delta^*, e) = P_m$, and that angle is set by the field fluxes $e$. After a disturbance the fluxes drift to new values, so the anchor moves. The generator is therefore a weight riding in a heavy sled that slides slowly along the floor, held near the sled's rest mark $\delta^*$ by the spring (Figure \ref{fig:sled}).

\begin{figure}[htbp]
\centering
\begin{tikzpicture}[scale=0.92, every node/.style={font=\small}]
  \begin{scope}[xscale=1.1]
  \draw[thick] (-0.3,0) -- (10.9,0);
  \foreach \x in {-0.1,0.3,...,10.7} \draw (\x,0) -- (\x-0.2,-0.2);
  \foreach \x in {1.2,7.8} \draw[thick,fill=white] (\x,0.075) circle (0.075);
  \draw[thick,fill=gray!12] (0.4,0.15) rectangle (8.6,0.32);
  \draw[thick,fill=gray!12] (0.4,0.32) rectangle (0.75,2.0);
  \draw[thick,fill=gray!12] (8.25,0.32) rectangle (8.6,2.0);
  \fill (5.75,0.15) -- (5.95,0.15) -- (5.85,0.32) -- cycle;
  \foreach \x in {5.55,6.85} \draw[thick,fill=white] (\x,0.385) circle (0.065);
  \draw[thick,fill=gray!40] (5.2,0.45) rectangle (7.2,1.75);
  \node[align=center,font=\footnotesize] at (6.2,1.1) {rotor\\mass $H/\omega_b$};
  \draw[decorate,decoration={zigzag,segment length=6pt,amplitude=3pt,pre length=4pt,post length=4pt}] (0.75,1.4) -- (5.2,1.4);
  \node[font=\footnotesize] at (3.0,1.72) {spring $K_s$};
  \draw (0.75,0.7) -- (4.0,0.7); \draw (4.0,0.56) -- (4.0,0.84);
  \draw (3.8,0.52) -- (4.7,0.52) -- (4.7,0.88) -- (3.8,0.88);
  \draw (4.7,0.7) -- (5.2,0.7);
  \node[font=\footnotesize] at (2.3,0.98) {damper windings};
  \draw[thick] (7.05,1.75) -- (7.05,2.45);
  \draw (7.05,2.45) -- (8.9,2.45); \draw (8.9,2.31) -- (8.9,2.59);
  \draw (8.7,2.27) -- (9.5,2.27) -- (9.5,2.63) -- (8.7,2.63); \draw (9.5,2.45) -- (10.3,2.45);
  \draw[thick] (10.3,0) -- (10.3,2.8);
  \foreach \y in {0.25,0.65,...,2.7} \draw (10.3,\y) -- (10.5,\y-0.2);
  \node[align=center,font=\footnotesize] at (8.6,3.25) {mechanical damping $D/\omega_b$\\(drag to the fixed synchronous frame)};
  \draw[<->,thick] (5.15,2.0) -- (6.45,2.0) node[midway,above,font=\footnotesize] {swings, 2.5\,Hz};
  \draw[->,thick] (0.9,2.4) -- (4.1,2.4) node[midway,above,align=center,font=\footnotesize] {sled $=$ transient fluxes $e'_q,e'_d$\\slides slowly ($\sim$1--2\,s)};
  \draw[dashed,gray] (5.85,0.15) -- (5.85,-0.6) node[anchor=north east,black,font=\footnotesize] {rest mark: anchor angle $\delta^*$};
  \draw[dashed,gray] (6.2,0.45) -- (6.2,-0.6) node[anchor=north west,black,font=\footnotesize] {rotor angle $\delta$};
  \draw[->,dashed,thick] (6.2,-1.3) to[out=200,in=-20] node[midway,below,font=\footnotesize] {reaction: through the stator currents the swing nudges the sled} (2.0,-1.3);
  \end{scope}
  \node at (5.7,-2.65) {(a)};

  \begin{scope}[yshift=-6.9cm, xshift=0.6cm]
    \fill[gray!18] (5.7,0) rectangle (9.5,3.0);
    \draw[thick] (0,0) rectangle (9.5,3.0);
    \foreach \x/\l in {0/0, 1.9/1, 3.8/2, 5.7/3, 7.6/4, 9.5/5} {
      \draw (\x,0) -- (\x,0.12);
      \node[below] at (\x,0) {\l};
    }
    \node[below=14pt] at (4.75,0) {$t\text{ (s)}$};
    \node[rotate=90, above=8pt] at (0,1.5) {angle (schematic)};
    \draw[thick, dashed, domain=0:9.5, samples=80] plot (\x, {2.5 - 1.1*exp(-0.28*\x)});
    \draw[very thick, domain=0:9.5, samples=250] plot (\x, {2.5 - 1.1*exp(-0.28*\x) - 1.2*exp(-2.2*\x)*cos(470*\x)});
    \node[right] at (1.5,2.5) {swing: the fast fibre, gone in $\approx 1\text{ s}$};
    \draw[->, >=Stealth] (1.45,2.45) -- (0.85,1.95);
    \node at (7.6,1.6) {window of \cite{evangelou2024}};
    \node[align=left, font=\footnotesize] at (3.0,0.65) {weight rides with the sled;\\lag $\sim 10^{-4}\text{ rad}$, invisible here};
    \node[align=left, font=\footnotesize] at (7.3,0.55) {--- rotor angle $\delta$ (weight)\\- - anchor angle $\delta^*$ (sled)};
    \node at (4.75,-1.35) {(b)};
  \end{scope}
\end{tikzpicture}
\caption{The swing as a weight riding in a sled: (a) mechanical analogue
of \eqref{eq:swing_mech}; (b) schematic time course, with the window
$[3, 5]\text{ s}$ opening after the swing has died out.}
\label{fig:sled}
\end{figure}

The two motions run on different clocks. The weight swings at about $2.5\text{ Hz}$, and the swing dies out within roughly a second. Most of that damping comes from the damper windings, a much stronger dashpot between the weight and the sled; mechanical friction $D$ supplies less than $3\%$ of it. The sled moves on the time scale of the transient fluxes, one to two seconds per $e$-fold. Once the swing has died out, the weight simply rides along with the sled, held close to $\delta^*$. A record that starts after that moment shows the motion of the sled, and the motion of the sled does not involve the weight's mass ($H$) or friction ($D$). $H$ and $D$ still matter. The window simply opens after the motion they govern has finished.

A gravity pendulum hanging from a moving support is the more familiar image, but it gets one thing wrong. Its restoring force, the weight $mg$, grows with its mass, so the mass cancels from the motion and from the lag behind the support. The generator's restoring force comes from the network through $K_s$ and does not scale with the inertia. Only a weight held by a spring keeps $H$ where the swing equation puts it.

Two faint traces of the swing do survive into a late record.
\begin{enumerate}
\item[(i)] \emph{Tilt: the weight lags behind the sled.} A weight riding in a moving sled does not sit exactly at the rest mark, because its inertia and the drag $D$ both act in the fixed frame, not in the sled's frame. Balancing the forces in \eqref{eq:swing_mech} during the slow motion gives, to leading order,
\[
\delta - \delta^* \approx -\frac{1}{K_s}\left(\frac{D}{\omega_b}\dot{\delta}^* + \frac{H}{\omega_b}\ddot{\delta}^*\right):
\]
the ground drag $D$, which resists the sled's speed, and the inertia of the weight, which resists the sled's acceleration. The damper windings act inside the electrical link between weight and sled; they tilt the slow manifold through the lag of the sub-transient fluxes (Proposition~\ref{prop:reentry}) rather than through this force balance. Both terms carry the factor $1/\omega_b$, and both are small because the sled moves slowly; on the late window the lag is of order $10^{-4}\text{ rad}$. The friction term is first order in the slowness of the sled and the mass term second order, which is the order in which $D$ and $H$ reappear in Section \ref{sec:late}.
\item[(ii)] \emph{Memory: the swing has nudged the sled.} The coupling runs both ways. The swinging rotor changes the stator currents, and the currents act back on the fluxes through armature reaction, so while the weight swings it jostles the sled. When the swing has died out the sled has been displaced by a small net amount, and that amount depends on how the swing went, hence on $H$, $D$ and the damper time constants. The displacement is then carried forward by the sled's own slow motion. To a late observer it looks exactly like a sled that started from a slightly different place, and other parameters, the reactances for instance, can produce that just as well. This trace is confounded.
\end{enumerate}

Tilt and memory are the whole mechanism of sloppiness for the mechanical parameters. The same picture predicts how a model-reduction method behaves when it walks through parameter space along the direction in which the record changes least. Shrinking the weight's mass or friction, or the time constants of the damper windings, only shrinks the small lag and the small nudge, so the faint traces fade in proportion to the parameter. A walk that drives one of these parameters to zero therefore ends after a distance equal to how strongly the parameter is felt now. Once all four have been driven to zero, it ends at a model in which the weight is rigidly fixed to the sled, $\delta = \delta^*$: electrical power always equals mechanical power. Had the record started while the weight was still swinging, the walk would have ended somewhere else.

The last ingredient is measurement noise. A parameter is identifiable when changing it within its range of uncertainty changes the record by more than the measurement error. The model fixes the size of the change; only the instruments fix the size of the error. Real instruments are lopsided. A phasor measurement unit (PMU) reports angle to about $1\%$ total vector error but frequency to $5\text{ mHz}$, which in per-unit terms is roughly a hundred times finer. That ratio decides several of the results in Section \ref{sec:noise}.

In what follows we make these pictures precise, show that elementary tools reproduce the conclusions information geometry reaches by geodesic integration, and evaluate the results under realistic synchrophasor noise.

\subsection{The information-geometric approach}
A model is structurally identifiable when the map from parameters to noise-free outputs is injective, and practically identifiable when the parameters can be estimated to useful precision from finite, noisy records \cite{saccomani2003}. Multi-parameter models in many fields share a characteristic structure, termed sloppiness: the eigenvalues of the Fisher information matrix (FIM) spread over many decades, roughly evenly on a logarithmic scale, so that a few parameter combinations are well determined and most are not \cite{gutenkunst2007,transtrum2010,transtrum2011}. The FIM is the metric that the model manifold, the image of parameter space in data space, induces on the parameters \cite{transtrum2010,transtrum2011}. Information geometry turns the local eigenvalue picture into a global one. The model manifold is a bounded hyper-ribbon whose boundaries correspond to physically interpretable limiting approximations. The manifold boundary approximation method (MBAM) \cite{transtrum2014} reduces a model by integrating a geodesic along the sloppiest direction until the metric degenerates, recognising the parameter limit that has been reached, taking it, and repeating. In power systems these ideas have been used for model reduction \cite{transtrum2017} and model verification \cite{transtrum2018}.

The work of \cite{evangelou2024} (henceforth called ESKT) studied the infinite-bus generator, a sixth-order model with eleven independently variable parameters observed with full state information on the late window $t \in [3, 5]\text{ s}$ after a disturbance, with these tools. The FIM spans eleven decades, and the authors describe the model as effectively six-dimensional at a cut-off of $10^{-2}$, with the identifiable directions aligned with six bare parameters. (In our reproduction of their spectrum the sixth eigenvalue, carried by $\delta T_q$, is $9.3 \times 10^{-3}$, just below the cut-off, so a strict count gives five; Table \ref{tab:fim_spec}.) MBAM removes the other five in the order $D \to 0$, $H \to 0$, $T_{d0}'' \to 0$, $T_{q0}'' \to 0$ and $\delta x_1 \to 0$. Output diffusion maps and geometric harmonics on $10^4$ sampled trajectories report the same six parameters. The authors note that both routes agree with classical singular-perturbation analysis, find the unidentifiability of $H$ and $D$ surprising since such parameters are usually associated with the slowest dynamics, attribute it to the rotor speed starting close to equilibrium, and conjecture two remedies: start farther from equilibrium, or use a disturbance that moves the operating point.

The agreement with singular-perturbation analysis is the starting point of this paper. It is reported in \cite{evangelou2024} as a finding. Nothing in the information-geometric construction explains why it should hold, when it should fail, or what it implies beyond the single experiment analysed. MBAM is deliberately structure-agnostic. On models with no exploitable structure this is its strength. On a model with a time-scale hierarchy it has a cost: each step requires second directional derivatives and numerical integration of the geodesic equation; the limit reached must be identified after the fact; and the whole computation describes one fixed choice of observation window, initial condition and operating point.

\subsection{The state-space approach of this paper}
We work from the geometry of the dynamics rather than from that of the parameter-to-data map. Synchronous machines, and converters with nested current, voltage and synchronisation loops, are built from processes acting on separated time scales. In the limit of wide separation such systems possess a normally attracting invariant manifold (NAIM): a slow manifold, written as a graph $y = h(x)$ of the fast variables over the slow ones, that attracts nearby trajectories exponentially along fast fibres \cite{fenichel1979,jones1995,kokotovic1986}. At the finite separation of a real machine we use this structure as an approximation, computed by a convergent iteration, and verify its consequences numerically against the full model. We show that the information a late record carries about any parameter is fixed by this geometry. The argument rests on one algebraic equation and on first-order sensitivities, and it has four parts.

\paragraph{Zeroth-order invisibility.} The zeroth-order graph $h_0$ solves the algebraic equation obtained by setting the fast vector field to zero. For the generator this equation does not contain $H$, $D$, $T_{d0}''$ or $T_{q0}''$ (Proposition \ref{prop:zeroth}). The reduced slow model is therefore exactly independent of these four parameters, and their late-window sensitivity columns vanish at leading order.

\paragraph{Three residual channels.} Once the fibre transient has decayed, a parameter that enters only the fast equations can reach the record through three channels only (Lemma \ref{lem:channels}): a fibre residual, which decays at the fast rate; a memory channel, the shift of the point at which the trajectory lands on the manifold, thereafter carried by the slow flow (Section \ref{subsec:memory} explains this channel in detail); and a tilt, the higher-order corrections $h_1, h_2, \dots$ to the graph, which are suppressed by the time-scale ratio.

\paragraph{Confounding.} The memory directions lie almost entirely in the span of the sensitivity columns of the slow parameters; more than $99.6\%$ of their norm is captured there. When the slow parameters are profiled out, the memory information disappears with them and only the small tilt remains. This, rather than any weakness of the swing dynamics, is what makes $H$ and $D$ sloppy.

\paragraph{Boundaries without geodesics.} Because the late-window prediction is affine to first order in each fast-only parameter, the distance along the model manifold from the current prediction to the boundary where that parameter vanishes equals its current profiled speed in logarithmic coordinates (Theorem \ref{thm:boundary}). When the sloppiest eigenvector is concentrated on one parameter, it points along the compensated direction of that parameter; the ranking of the distances matches the order in which MBAM meets the boundaries; and the model reached after the four limits is the zeroth-order slow-manifold model on $h_0$.

We then evaluate the Fisher analysis in a physical metric fixed by synchrophasor accuracy limits \cite{ieee_c37118} and connect the analytical spectrum directly to the covariance of sampled trajectory ensembles. Under PMU-grade noise the late window identifies five parameters rather than six, while the early window $[0, 2]\text{ s}$ identifies nine, with $H$ the best determined. We also show that in the linearised regime the covariance of whitened trajectory ensembles has a spiked-covariance form governed by the prior-scaled Fisher matrix (Proposition \ref{prop:spiked}). This explains why analyses of sampled trajectories reproduce the Fisher results, and why per-feature $[0, 1]$ scaling, which amounts to an implicit noise model, can lift physically unidentifiable directions to the resolution limit.

\subsection{What the NAIM analysis adds}
For systems with time-scale separation, and on the boundaries that come from that separation, the NAIM analysis recovers what MBAM finds and adds to it. It gives the mechanism behind the sloppiness of $H$ and $D$, zeroth-order invisibility plus memory confounding, and so explains the agreement with singular-perturbation theory that \cite{evangelou2024} reports as an observation. It predicts the null set, the boundary distances and the expected order of the boundaries from the invariance equation and first-order sensitivities, with no geodesic integration and no second derivatives. It also writes down the reduced model. The terminal model is the slow dynamics on $h_0$, and the iterated graphs $h_1, h_2, \dots$ give a hierarchy of corrected models with explicit error orders, whereas MBAM delivers only the terminal model and leaves its surviving parameter combinations to be interpreted afterwards.

The analysis treats the experiment as a variable. The FIM and MBAM describe one fixed experiment and must be recomputed when it changes, while the NAIM analysis predicts how identifiability depends on it. A larger initial speed deviation or a new operating point changes only the two memory numbers and the slow amplitude that feeds the tilt, so neither restores $H$ or $D$ on the late window. Moving the window start into the fibre transient restores $H$, which there becomes the best-determined parameter. $D$, whose effect on this machine is overshadowed by the damper windings, also needs a larger initial speed deviation and accurate frequency measurement. Finally, the argument depends on where parameters enter the fast and slow equations, not on a particular parameter point, so it should carry over to other components with a time-scale hierarchy, such as grid-forming and grid-following converters \cite{singh2026}.

\subsection{Scope and organisation}
These gains depend on a usable spectral gap, and the information-geometric tools keep a role that the NAIM analysis does not take over. Three cases fall outside it:
\begin{itemize}
\item \emph{Regular near-degeneracies among the slow parameters.} The fifth MBAM limit, $\delta x_1 \to 0$, is of this kind and is not predicted by the time-scale structure.
\item \emph{The absolute identifiability threshold.} The NAIM analysis fixes the ordering and the structural origin of the small eigenvalues, which do not depend on the noise level. Whether a parameter is identifiable from a given record still depends on the noise and is decided by the Fisher metric. Under PMU-grade noise \cite{ieee_c37118} the ``identifiable six'' of \cite{evangelou2024} no longer holds (Section \ref{sec:noise}).
\item \emph{Regimes without a clear gap}, where eigenvalues cluster or a bifurcation is approached. There Fenichel theory provides no normally attracting manifold, and MBAM, or a structure-agnostic data-driven method, is required.
\end{itemize}

Section \ref{sec:model} states the generator model, reproduces the published Fisher spectrum after correcting two transcription errors, and sets up the fast-slow split. Section \ref{sec:late} develops the zeroth-order invisibility, the three residual channels, a self-contained explanation of the memory channel, and memory confounding. Section \ref{sec:boundaries} proves the boundary theorem, compares it with the MBAM sequence of \cite{evangelou2024}, and examines how the boundaries move with the experiment. Section \ref{sec:noise} repeats the analysis under PMU-grade measurement noise and establishes the analytical link to sampled trajectory ensembles. Section \ref{sec:discussion} compares our findings with \cite{evangelou2024}, states the limitations and concludes. Computational details are collected in Appendix \ref{app:comp}.

\section{The generator model and its fast-slow split}\label{sec:model}

\subsection{Model}
The model is Eqs.~(4)--(14) of \cite{evangelou2024}, in the standard form of \cite{sauer2017}, with state $z = (\delta, \omega, e_q', e_d', e_q'', e_d'')$:
\begin{equation}
\begin{aligned}
\dot{\delta} &= \omega_b(\omega - 1)\\
\dot{\omega} &= \frac{1}{H}(P_m - P_g - D(\omega - 1)) \\
T_{d0}'\dot{e}_q' &= -e_q' - (x_d - x_d')i_d + v_{f0}\\
T_{q0}'\dot{e}_d' &= -e_d' + (x_q - x_q')i_q \\
T_{d0}''\dot{e}_q'' &= -e_q'' + e_q' - (x_d' - x_d'')i_d\\
T_{q0}''\dot{e}_d'' &= -e_d'' + e_d' + (x_q' - x_q'')i_q\\
\end{aligned}
\label{eq:gen_model}
\end{equation}
with
\[
v_d = V\sin(\delta - \vartheta), \quad v_q = V\cos(\delta - \vartheta), \quad i_d = \frac{e_q'' - v_q}{x_d''}, \quad i_q = \frac{v_d - e_d''}{x_q''}, \quad P_g = v_d i_d + v_q i_q.
\]
The constants are $\omega_b = 120\pi\text{ rad/s}$, $v_{f0} = 4.2$, $P_m = 0.7$, $V = 1.09$ and $\vartheta = 0$. The post-disturbance initial state is $z(0) = (0.5, 0.98, 2.13, 0.02, 1.93, 0.02)$. The eleven independent parameters,
\[
\theta = (H, D, \delta x_1, \delta x_2, \delta x_3, \delta x_4, x_d'', \delta T_d, \delta T_q, T_{d0}'', T_{q0}''),
\]
are chosen so that the ordering constraints on the reactances and time constants hold under independent perturbation. Their nominal values are
\[
(2.53, 0.5, 0.12, 2.02, 1.932, 0.448, 0.48, 4.69, 1.29, 0.06, 0.21).
\]
Here $\delta x_1 = x_d - x_q$, $\delta x_2 = x_q - x_q'$, $\delta x_3 = x_q' - x_d'$, $\delta x_4 = x_d' - x_q''$, $\delta T_d = T_{d0}' - T_{d0}''$, $\delta T_q = T_{q0}' - T_{q0}''$, and $x_q'' = x_d''$. All parameters are varied in logarithmic coordinates.

\subsection{Transcription corrections}
Reproducing the results of ESKT requires two corrections to their stated
equations; a third concerns only their reduced model.
\begin{itemize}
\item \emph{Speed reference.} The rotor speed is in per unit with reference
  $\omega_0 = 1$. The value $\omega_0 = 2\pi\cdot 60$ given in the text of ESKT
  is inconsistent with $\omega(0) = 0.98$ and $\omega_b = 120\pi$.
\item \emph{$q$-axis current.} The current is $i_q = (v_d - e_d'')/x_q''$, the
  standard form~\cite{sauer2017}; their Eq.~(13) has $e_q''$ in place of
  $e_d''$. Only the standard form reproduces their trajectories (Fig.~2 of
  ESKT).
\item \emph{Reduced model.} Their Eq.~(37) has $(x_d - x_q')\,i_q$ where
  Eq.~(7) has $(x_q - x_q')\,i_q$ 
\end{itemize}
\subsection{Sensitivities, Fisher information and reproduction}
Sensitivities $S(t) = \partial z(t)/\partial \ln \theta$ are computed from the forward sensitivity equations (Appendix \ref{app:comp}). With full-state observation every $0.02\text{ s}$ on $[3, 5]\text{ s}$ and unit noise, the FIM is $\mathcal{I} = \sum_t S(t)^\top S(t)$. Table \ref{tab:fim_spec} gives its spectrum. The sloppiest eigenvalue, $1.207 \times 10^{-9} = (3.47 \times 10^{-5})^2$, equals the value reported in \cite{evangelou2024} for the sloppiest ($D$) direction, and the participation pattern matches their Figure 3. Sensitivities are taken with respect to the log-parameters, $S = (\partial z/\partial \theta)\operatorname{diag}(\theta)$, and all Fisher quantities in Sections \ref{sec:model}--\ref{sec:boundaries} use the 101 samples of $[3, 5]\text{ s}$ at $0.02\text{ s}$.

\begin{table}[htbp]
\centering
\caption{Manifold boundaries on $[3,5]$\,s (unit noise), in the MBAM order of ESKT; symbols as in Definition~1 and Appendix~A.}
\begin{tabular}{cccccc}
\toprule
mode & eigenvalue & dominant & mode & eigenvalue & dominant \\
\midrule
1 & $1.4 \times 10^2$ & $\delta x_2, \delta x_3$ & 7 & $2.8 \times 10^{-4}$ & $\delta x_1$ \\
2 & $1.0 \times 10^1$ & $x_d'', \delta x_4$ & 8 & $2.9 \times 10^{-5}$ & $T_{q0}'', T_{d0}''$ \\
3 & $5.4$ & $\delta x_3, \delta x_2$ & 9 & $6.2 \times 10^{-6}$ & $T_{d0}'', T_{q0}''$ \\
4 & $3.8$ & $\delta x_4, x_d''$ & 10 & $7.5 \times 10^{-8}$ & $H$ \\
5 & $2.1 \times 10^{-1}$ & $\delta T_d$ & 11 & $1.2 \times 10^{-9}$ & $D$ \\
6 & $9.3 \times 10^{-3}$ & $\delta T_q$ & & & \\
\bottomrule
\end{tabular}
\label{tab:fim_spec}
\end{table}

\subsection{Time scales: the swing mode is a fast fibre}
Linearising at the post-disturbance equilibrium $z^* = (0.865, 1, 1.355, 0.344, 1.042, 0.748)$ gives the spectrum in Table \ref{tab:lin_spec}.

\begin{table}[htbp]
\centering
\caption{Spectrum of the linearisation at the post-disturbance equilibrium.}
\begin{tabular}{cccc}
\toprule
eigenvalue ($\text{s}^{-1}$) & time constant & frequency & dominant states \\
\midrule
$-0.536$ & $1.87\text{ s}$ & -- & $e_q'$ (with $\delta, e_q''$) \\
$-1.159$ & $0.86\text{ s}$ & -- & $e_d'$ \\
$-4.263 \pm 15.589j$ & $0.235\text{ s}$ & $2.48\text{ Hz}$ & $\delta, \omega$ (swing), $e_d'', e_q''$ \\
$-21.28$ & $0.047\text{ s}$ & -- & $e_d''$ \\
$-30.16$ & $0.033\text{ s}$ & -- & $e_q''$ \\
\bottomrule
\end{tabular}
\label{tab:lin_spec}
\end{table}

$H$ and $D$ appear only in the $\omega$-equation, whose dynamics is the swing mode. That mode decays $3.7$ times faster than the faster slow flux mode, and almost eight times faster than the slowest. It is damped mainly by the damper windings: the mechanical term $D/(2H) \approx 0.10\text{ s}^{-1}$ accounts for under $3\%$ of its decay rate of $4.26\text{ s}^{-1}$. By $t = 3\text{ s}$ the swing has decayed by a factor $e^{-4.26 \cdot 3} \approx 3 \times 10^{-6}$. In NAIM language, $(\delta, \omega)$ together with $(e_q'', e_d'')$ are fast fibre coordinates, and $(e_q', e_d')$ are the slow coordinates. The initial condition $\omega(0) = 0.98$ is not dynamically close to equilibrium either: it gives $\dot{\delta}(0) = \omega_b(\omega(0) - 1) = -7.5\text{ rad/s}$ and a large swing, which is over before the observation window opens.

\subsection{Fast-slow split and the invariance equation}
Write the model as
\[
\dot{x} = f(x, y; \theta), \qquad \dot{y} = G(x, y; \theta),
\]
with slow state $x = (e_q', e_d')$ and fast state $y = (\delta, \omega, e_q'', e_d'')$. A graph $y = h(x; \theta)$ is invariant exactly when it solves the invariance equation
\begin{equation}
G(x, h(x)) = \partial_x h(x) f(x, h(x)).
\label{eq:invariance}
\end{equation}
Fenichel's theory \cite{fenichel1979,jones1995} guarantees a solution near the critical manifold $G = 0$, together with an invariant fibration along which fast transients collapse onto it, for a family of systems in which the ratio of slow to fast rates tends to zero with the fast subsystem uniformly attracting. A single machine with a ratio of about $1/3.7$ is not such a limit, and we do not verify Fenichel's hypotheses analytically for it. We use the theory as a guide: the graph is computed by the iterated quasi-steady-state construction of Roussel and Fraser \cite{roussel1990},
\begin{equation}
G(x, h_0) = 0, \qquad G(x, h_{k+1}) = \partial_x h_k(x) f(x, h_k(x)), \quad k \ge 0,
\label{eq:fraser_iter}
\end{equation}
in which each step adds one order in the ratio of slow to fast rates, and every property we draw from it (graph error, slaving of the fast sensitivities, linearity of the profiled speeds) is checked numerically against the full model. The graph $h_0$ is the classical singular-perturbation reduction \cite{kokotovic1986}; $h_1$ adds the first-order tilt.

\begin{remark}[Relation to the Riccati equation]
The tilts in this paper are computed from the nonlinear invariance equation \eqref{eq:invariance} through the iteration \eqref{eq:fraser_iter}, not from a Riccati equation. At the linear level, however, the two are the same object. For a linear system $\dot{x} = A_{11}x + A_{12}y$, $\dot{y} = A_{21}x + A_{22}y$ and a linear graph $y = Lx$, equation \eqref{eq:invariance} becomes the non-symmetric algebraic Riccati equation $A_{21} + A_{22}L = LA_{11} + LA_{12}L$ of the decoupling transformation of singularly perturbed systems \cite{kokotovic1975,kokotovic1986}. Differentiating \eqref{eq:invariance} in $\theta$ gives a linear equation for the parameter sensitivity of the graph, $\partial h/\partial \theta$:
\[
G_y \frac{\partial h}{\partial \theta} + G_\theta = \partial_x\left(\frac{\partial h}{\partial \theta}\right)f + \partial_x h \left(f_y \frac{\partial h}{\partial \theta} + f_\theta\right).
\]
In the linear case, with $L_\theta = \partial L/\partial \theta$, this is the Sylvester equation
\[
(A_{22} - LA_{12})L_\theta - L_\theta(A_{11} + A_{12}L) = L(\partial_\theta A_{11} + \partial_\theta A_{12}L) - (\partial_\theta A_{21} + \partial_\theta A_{22}L),
\]
which has a unique solution whenever the spectra of the fast block $A_{22} - LA_{12}$ and the reduced slow block $A_{11} + A_{12}L$ are disjoint. The size of the spectral gap bounds the size of the solution, which is why the tilts of the fast-only parameters are suppressed by the time-scale ratio (Proposition \ref{prop:reentry}).
\end{remark}

\section{Why a late record cannot see the swing}\label{sec:late}

\subsection{Zeroth-order invisibility}

\begin{proposition}[Zeroth order]\label{prop:zeroth}
Wherever $G_y(x, h_0)$ is invertible, $h_0$ does not depend on $H$, $D$, $T_{d0}''$ or $T_{q0}''$. In the bare parameters, the zeroth-order reduced model
\[
\dot{x} = f(x, h_0(x)), \qquad y = h_0(x)
\]
is independent of all four when the bare time constants $T_{d0}'$ and $T_{q0}'$ are held fixed. Its FIM then has the four corresponding directions in its null space.
\end{proposition}

\begin{proof}
The four components of $G(x, h_0) = 0$ read as follows:
\begin{itemize}
\item $\omega_b(\omega - 1) = 0$, so $\omega = 1$.
\item $(P_m - P_g - D(\omega - 1))/H = 0$, which with $\omega = 1$ and $H > 0$ is $P_g = P_m$.
\item For $e_q''$ and $e_d''$, the numerators $-e_q'' + e_q' - (x_d' - x_d'')i_d$ and $-e_d'' + e_d' + (x_q' - x_q'')i_q$ vanish, since $T_{d0}'', T_{q0}'' > 0$.
\end{itemize}
None of these equations contains $H$, $D$, $T_{d0}''$ or $T_{q0}''$, so neither does their solution. The slow field $f$ contains $T_{d0}'$, $T_{q0}'$, the reactances, $v_{f0}$ and the currents, but none of the four.
\end{proof}

\begin{remark}[Choice of time-constant coordinates]\label{rem:coords}
In the independent coordinates $\theta$ of Section \ref{sec:model}, $T_{d0}''$ and $T_{q0}''$ still reach $f$ through $T_{d0}' = T_{d0}'' + \delta T_d$ and $T_{q0}' = T_{q0}'' + \delta T_q$. Let $\theta'$ be the coordinates in which $T_{d0}'$ and $T_{q0}'$ replace $\delta T_d$ and $\delta T_q$. In log-coordinates the sensitivity columns are related by
\[
\left.\frac{\partial y}{\partial \ln T_{d0}''}\right|_{\delta T_d} = \left.\frac{\partial y}{\partial \ln T_{d0}''}\right|_{T_{d0}'} + \frac{T_{d0}''}{\delta T_d}\frac{\partial y}{\partial \ln \delta T_d}, \qquad \frac{\partial y}{\partial \ln \delta T_d} = \frac{\delta T_d}{T_{d0}'}\frac{\partial y}{\partial \ln T_{d0}'},
\]
and similarly on the $q$-axis. The two coordinate systems therefore differ only by a term in the span of the remaining columns, and the profiled speed of $T_{d0}''$ (and of $T_{q0}''$), which projects that span out, is the same in both (agreeing numerically on $[3, 5]\text{ s}$ to ten digits). The FIM spectrum itself is not invariant: in the coordinates $\theta'$ the sixth eigenvalue is $1.26 \times 10^{-2}$ instead of $9.3 \times 10^{-3}$. A count of eigenvalues above a sharp cut-off is a coordinate-dependent quantity, which is one reason we rely on profiled quantities.
\end{remark}

\begin{remark}[Four of the five MBAM limits]\label{rem:four_mbam}
MBAM \cite{transtrum2014,transtrum2017} removes $D$, $H$, $T_{d0}''$, $T_{q0}''$ and $\delta x_1$, one geodesic at a time. Proposition \ref{prop:zeroth} obtains the first four in one step: they are exactly the parameters invisible to the zeroth-order NAIM. The fifth is different in kind. The parameter $\delta x_1 = x_d - x_q$ enters the slow field through $x_d - x_d' = \delta x_1 + \delta x_2 + \delta x_3$, where it is a $3\%$ addend to a sum dominated by $\delta x_2 + \delta x_3$. Its sloppiness in logarithmic coordinates is a regular, not a singular, near-degeneracy, and Proposition \ref{prop:zeroth} does not predict it.
\end{remark}

\begin{proposition}[Order of re-entry]\label{prop:reentry}
Let $L_f$ denote the derivative along the zeroth-order slow flow. In the iteration \eqref{eq:fraser_iter}:
\begin{enumerate}
\item[(a)] $h_1$ depends on $D$, $T_{d0}''$ and $T_{q0}''$, and not on $H$. Explicitly,
\begin{align*}
\omega_1 &= 1 + \frac{1}{\omega_b}L_f \delta_0, & P_g(\delta_1, e_1'') &= P_m - \frac{D}{\omega_b}L_f \delta_0, \\
-e_q'' + e_q' - (x_d' - x_d'')i_d &= T_{d0}'' L_f e_{q,0}'', & -e_d'' + e_d' + (x_q' - x_q'')i_q &= T_{q0}'' L_f e_{d,0}''.
\end{align*}
\item[(b)] $H$ first appears in $h_2$, through
\[
P_g = P_m - D(\omega - 1) - \frac{H}{\omega_b}L_f^2 \delta_0,
\]
that is, with coefficient $H/\omega_b$ multiplying the second derivative of $\delta$ along the slow flow.
\end{enumerate}
\end{proposition}

\begin{proof}
(a) The right side of \eqref{eq:fraser_iter} for $k=0$ is $\partial_x h_0 f$, whose components are the slow-flow derivatives of $h_0$. Its $\omega$-component is $L_f \omega_0 = L_f 1 = 0$. The four equations $G(x, h_1) = \partial_x h_0 f$ are therefore those displayed: the $\delta$-equation fixes $\omega_1$; the $\omega$-equation is $(P_m - P_g - D(\omega_1 - 1))/H = 0$, in which $H$ cancels; and the $e''$-equations carry $T_{d0}''$ and $T_{q0}''$ as multipliers of nonzero right sides.
(b) At $k=1$ the $\omega$-component of $\partial_x h_1 f$ is $L_f \omega_1 = \omega_b^{-1}L_f^2 \delta_0 \neq 0$. The $\omega$-equation becomes $P_m - P_g - D(\omega - 1) = H L_f \omega_1$, and $H$ no longer cancels.
\end{proof}

\begin{remark}[Why $\omega$ looks constant, and the ranking]\label{rem:ranking}
On the manifold, $\omega - 1 = \dot{\delta}/\omega_b$ to first order. At $t = 3\text{ s}$, $\dot{\delta} \approx 0.05\text{ rad/s}$, so $\omega - 1 \approx 1.3 \times 10^{-4}$ whatever $\omega(0)$ was. The near-constancy of $\omega$ is this slaving, not a property of the initial condition. The same factor $1/\omega_b = 2.65 \times 10^{-3}$ multiplies the tilt of both swing parameters, which is why $H$ and $D$ sit below $T_{d0}''$ and $T_{q0}''$ in the sloppy hierarchy. Formal order alone would put $H$ (second order) below $D$ (first order). The published hierarchy has the opposite order, and the coefficients explain why: the $\delta$-tilts are in the ratio $|H\ddot{\delta}|/|D\dot{\delta}| \approx (H/D) \cdot 0.54 = 2.7$, taking $\ddot{\delta} \approx -0.54\dot{\delta}$ on the slowest mode; the computed ratio at $t = 3\text{ s}$ is $1.32 \times 10^{-4} / 4.79 \times 10^{-5} = 2.75$.
\end{remark}

\begin{figure}[htbp]
\centering
\incfig{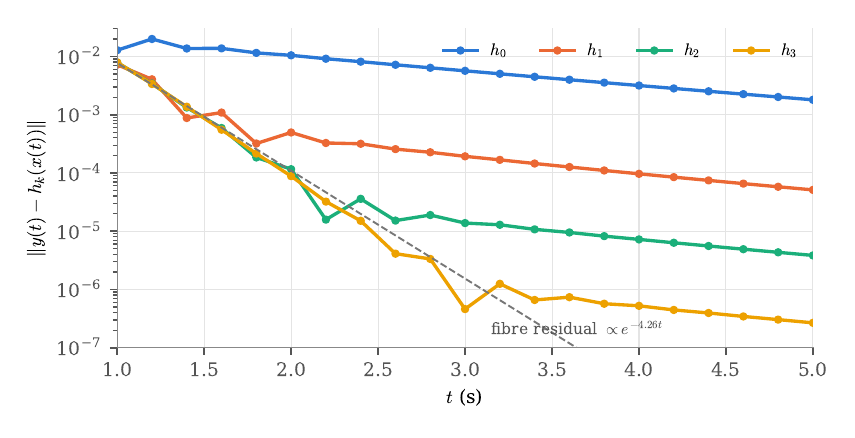}
\caption{Distance from the true trajectory to the iterated NAIM graphs
$h_k$, $k = 0,\dots,3$.}
\label{fig:graph_err}
\end{figure}

\paragraph{Numerical verification.} We computed $h_0, \dots, h_3$ and their parameter derivatives along the true trajectory in 130-digit arithmetic (Appendix \ref{app:comp}). We find $\partial h_0/\partial \ln \theta = 0$ exactly for $\theta \in \{H, D, T_{d0}'', T_{q0}''\}$, and $\partial h_1/\partial \ln H = 0$ exactly, at every point tested on $t \in [1, 5]\text{ s}$. The remaining derivatives are small: $\max_t |\partial h_1/\partial \ln D| = 1.3 \times 10^{-4}$, $|\partial h_1/\partial \ln T_{d0}''| = 6.5 \times 10^{-3}$, $|\partial h_1/\partial \ln T_{q0}''| = 8.0 \times 10^{-3}$ and $|\partial h_2/\partial \ln H| = 3.5 \times 10^{-4}$. The graph error $\|y(t) - h_k(x(t))\|$ at $t = 3\text{ s}$ is $5.7 \times 10^{-3}$, $1.9 \times 10^{-4}$, $1.4 \times 10^{-5}$ and $4.6 \times 10^{-7}$ for $k = 0, 1, 2, 3$, a gain of about a factor 25 per order (Figure \ref{fig:graph_err}). Before $t \approx 2.5\text{ s}$ all corrected graphs hit the same floor, the fibre residual, which decays as $e^{-4.26t}$.

\subsection{Three residual channels}

\begin{lemma}[Three channels]\label{lem:channels}
Assume that the trajectory is attracted to a NAIM $y = h(x; \theta)$ of dimension $m$ with a smooth invariant fibration whose fibres contract at rate at least $\lambda_f$ \cite{fenichel1979,jones1995}. For a parameter $\theta$ that enters only the fast equations, and for $t \ge t_1$,
\begin{align}
\frac{\partial y}{\partial \theta}(t) &= \underbrace{\frac{\partial h}{\partial \theta}}_{\text{tilt}} + \partial_x h \frac{\partial x}{\partial \theta}(t) + \underbrace{O(e^{-\lambda_f t})}_{\text{fibre residual}}, \label{eq:chan_y} \\
\frac{\partial x}{\partial \theta}(t) &= \underbrace{\Phi_s(t, t_1) c_\theta}_{\text{memory, } c_\theta \in \mathbb{R}^m} + \int_{t_1}^t \Phi_s(t, s) f_y \frac{\partial h}{\partial \theta}\,ds + O(e^{-\lambda_f t_1}), \label{eq:chan_x}
\end{align}
where $\Phi_s$ is the transition matrix of the reduced flow linearised along the trajectory and $c_\theta = \partial x(t_1)/\partial \theta$. In particular, every perturbation of the initial condition produces, for $t \ge t_1$, an output change in the $m$-dimensional span $\{\partial_x h \Phi_s(t, t_1) e_i\}_{i=1}^m$, up to $O(e^{-\lambda_f t})$.
\end{lemma}

\begin{proof}[Sketch]
Write $y = h(x; \theta) + r$ with $r$ the fibre offset and $|r| = O(e^{-\lambda_f t})$ by normal attraction, and differentiate in $\theta$ to obtain \eqref{eq:chan_y}. Substituting into $\dot{x} = f(x, h+r)$ and differentiating gives a linear equation for $\partial x/\partial \theta$, forced by $f_y \partial h/\partial \theta$ plus $O(e^{-\lambda_f t})$; variation of constants from $t_1$ gives \eqref{eq:chan_x}. The final statement is the asymptotic-phase property of the fibration: an initial perturbation moves only the base point on the manifold, which is an $m$-dimensional quantity.
\end{proof}

\begin{figure}[htbp]
\centering
\incfig{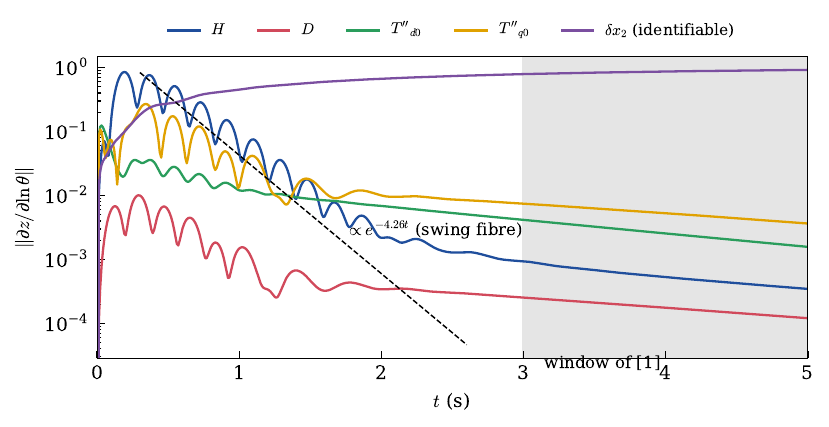}
\caption{Norm of the state sensitivity to each log-parameter; $\delta x_2$
is shown for comparison. Shaded: the window $[3, 5]\text{ s}$ of ESKT.}
\label{fig:sens_norm}
\end{figure}

The lemma separates three channels for a fast-only parameter: the fibre residual, which is exponentially gone; the memory, which consists of $m = 2$ numbers $c_\theta$ fixed during the transient; and the tilt. Figure \ref{fig:sens_norm} shows the transition from the first regime to the other two. While the trajectory collapses onto the NAIM, the sensitivities to $H$, $D$, $T_{d0}''$ and $T_{q0}''$ decay with the swing envelope; afterwards they settle onto a slow tail carried by memory and tilt. $D$ reaches its tail by $t \approx 1.5\text{ s}$, $H$ only by $t \approx 2.5\text{--}3\text{ s}$, which is precisely where the window of \cite{evangelou2024} begins.

\subsection{What the memory channel is}\label{subsec:memory}
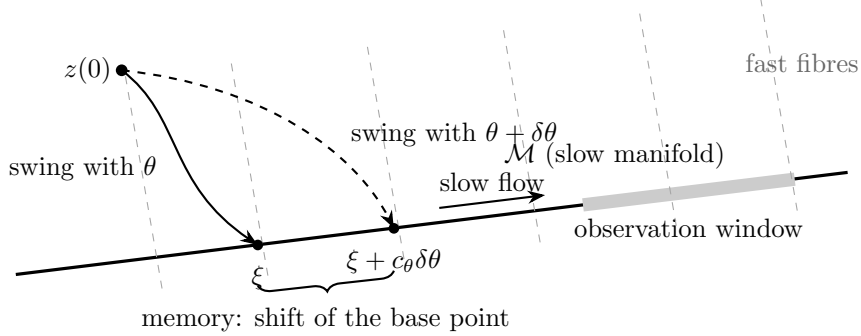
\begin{figure}[htbp]
\centering
\begin{tikzpicture}[scale=1.0, every node/.style={font=\small}]
  \draw[very thick] (0,0) -- (11,1.35) node[pos=0.72, above=8pt] {$\mathcal{M}$ (slow manifold)};
  \draw[line width=5pt, gray!40] (7.5,0.92) -- (10.3,1.26);
  \node[below] at (8.9,0.88) {observation window};
  \foreach \x/\y in {1.8/0.22, 3.2/0.39, 5.0/0.61, 6.8/0.83, 8.6/1.05, 10.2/1.25} {
    \draw[dashed, gray!70] (\x-0.35,\y+2.4) -- (\x+0.15,\y-0.5);
  }
  \node[gray!80!black] at (10.4,2.8) {fast fibres};
  \fill (1.4,2.7) circle (2.2pt) node[left] {$z(0)$};
  \draw[thick, ->, >=Stealth] (1.4,2.7) .. controls (2.2,2.1) and (2.0,1.2) .. (3.2,0.39);
  \node[left] at (2.0,1.4) {swing with $\theta$};
  \fill (3.2,0.39) circle (2pt) node[below=4pt] {$\xi$};
  \draw[thick, dashed, ->, >=Stealth] (1.4,2.7) .. controls (3.3,2.5) and (4.5,1.6) .. (5.0,0.61);
  \node[right] at (4.3,1.85) {swing with $\theta + \delta\theta$};
  \fill (5.0,0.61) circle (2pt) node[below=4pt] {$\xi + c_\theta \delta\theta$};
  \draw[->, >=Stealth, thick] (5.6,0.88) -- (7.0,1.05) node[midway, above] {slow flow};
  \draw[decorate, decoration={brace, mirror, amplitude=5pt}, thick] (3.2,-0.15) -- (5.0,0.05) node[midway, below=7pt] {memory: shift of the base point};
\end{tikzpicture}
\caption{The memory channel ($\mathcal{M}$ drawn one-dimensional; $m = 2$
for the generator).}
\label{fig:memory}
\end{figure}

Of the three channels, memory is the one that decides identifiability here, and it is the least familiar. Picture a ball released on the steep wall of a valley whose floor slopes gently downhill. The ball first rolls quickly down the wall, perhaps rocking from side to side, and then travels slowly along the floor. The wall is a fast fibre, the floor is the slow manifold. Once the ball is on the floor, its future depends only on where along the floor it arrived; the steepness of the wall, the friction on it and the details of the rocking leave no other trace. If an observer watches only the slow part of the journey, everything that happened on the wall is compressed into one fact: the arrival point. The record remembers the fast transient through that point alone. This is the memory channel (Figure \ref{fig:memory}).

Fenichel's theory \cite{fenichel1979,jones1995} attaches to every state $z$ near the manifold a base point $\xi = \pi_\theta(z)$ on the manifold: the point whose fibre passes through $z$. Its defining property, called asymptotic phase, is that the solution through $z$ and the solution on the manifold through $\xi$ converge at the fast rate,
\[
\|z(t) - Z_s(t; \xi)\| \le C e^{-\lambda_f t}, \qquad \xi = \pi_\theta(z(0)),
\]
where $Z_s(t; \xi) = (X_s(t; \xi), h(X_s(t; \xi)))$ is the slow solution started at $\xi$. Once the fibre residual is negligible, the record is therefore a function of only two things: the base point $\xi$, which is two numbers, and the slow dynamics on the manifold, which includes its graph $h$. A parameter $\theta$ that enters only the fast equations can act through both: its effect on $\xi$, $c_\theta = \partial\xi/\partial\theta \in \mathbb{R}^m$, is the memory; its effect on $h$ is the tilt. Because the base point has only $m = 2$ coordinates, the memory of all four fast parameters reaches the late window through the same two-dimensional set of output directions $\{\partial_x h \Phi_s(t, t_1)e_i\}$, which the slow parameters also span.

\subsection{Memory confounding}
Three consequences of Lemma \ref{lem:channels} can be checked directly on $[3, 5]\text{ s}$:
\begin{enumerate}
\item \emph{The memory channel is two-dimensional.} The singular values of the initial-condition sensitivity $\partial z(t)/\partial z_0$, stacked over the samples of $[3, 5]\text{ s}$ at $0.02\text{ s}$, are $1.04$, $0.20$, $3.7 \times 10^{-5}$, $3.1 \times 10^{-6}$, $5 \times 10^{-17}$, $2 \times 10^{-17}$. Two directions survive; the next two are the swing residual ($\sim e^{-4.26 \cdot 3}$), and the last two the sub-transient residual.
\item \emph{The identifiable parameters span that channel.} Projecting the two memory directions onto the span of the sensitivity columns of the seven non-fast parameters captures $99.94\%$ and $99.61\%$ of their norm. Profiling out those seven parameters reduces the information on $H$ from $3.5 \times 10^{-5}$ to $1.5 \times 10^{-7}$, and on $D$ from $3.4 \times 10^{-6}$ to $8.1 \times 10^{-9}$.
\item \emph{The fast rows are slaved.} On $[3, 5]\text{ s}$, identity \eqref{eq:chan_y} with $h = h_2$, using the exact full-model $\partial x/\partial \theta$, reproduces the full-model fast-state sensitivities to a relative error of at most $4.8 \times 10^{-2}$ for $H$, $2.0 \times 10^{-3}$ for $D$, $2.5 \times 10^{-2}$ for $T_{d0}''$ and $6.3 \times 10^{-3}$ for $T_{q0}''$.
\end{enumerate}

\section{Manifold boundaries without geodesics}\label{sec:boundaries}

\subsection{The boundary theorem}
MBAM \cite{transtrum2014} works on the model manifold $\mathcal{M}_W = \{y_W(\theta)\}$ with the Fisher matrix $\mathcal{I} = J^\top J$ as metric, where $J = \partial y_W/\partial \ln \theta$, integrating the geodesic equation $\ddot{\theta} = -(J^\top J)^{-1}J^\top A_{vv}$ along the sloppiest direction until the metric degenerates. We show that for the parameters in the zeroth-order null set, the slow-manifold structure supplies the boundary locations, their distances, and the reduced model without integrating a geodesic.

\begin{definition}[Compensated curve and profiled speed]\label{def:profiled}
Let $F$ be the set of free log-parameters and $j \in F$. The profiled speed is
\[
p_j(\theta) = \left(\mathcal{I}_{jj} - \mathcal{I}_{jO}\mathcal{I}_{OO}^{-1}\mathcal{I}_{Oj}\right)^{1/2} = \left[(\mathcal{I}^{-1})_{jj}\right]^{-1/2}, \qquad O = F \setminus \{j\}.
\]
It is the data-space speed of the compensated curve along which $\ln \theta_j$ changes while the other free parameters move, to first order, so as to keep the prediction as close as possible to the current one. Its length from the current point to the limit $\theta_j \to 0$ is
\[
d_j = \int_{-\infty}^{\ln \theta_j} p_j\,d(\ln \theta_j') = \int_0^{\theta_j} \frac{p_j(\theta')}{\theta'}\,d\theta'.
\]
\end{definition}

\begin{theorem}[Boundaries from the slow manifold]\label{thm:boundary}
Let $\theta_j$ belong to the zeroth-order null set $\{H, D, T_{d0}'', T_{q0}''\}$ of Proposition \ref{prop:zeroth}, and let the window $W = [t_a, t_b]$ start after the fibre transient. Assume:
\begin{enumerate}
\item[(B1)] the fibres keep a contraction rate $\lambda_f > 0$ as $\theta_j \to 0$, and the base point of the fibre through the initial condition depends smoothly on $\theta_j$ up to $\theta_j = 0$;
\item[(B2)] $\theta_j$ enters the iterated graphs only as the coefficient of its first tilt, $h_k = h_k|_{\theta_j=0} + \theta_j \partial_{\theta_j} h_k + O(\theta_j^2)$, as in Proposition \ref{prop:reentry};
\item[(B3)] along the compensated curve down to $\theta_j = 0$, the sensitivity columns of the other free parameters remain continuous and of full rank, and the projected coefficient $\Pi_j b_j$ defined below is nonzero.
\end{enumerate}
Then the following hold:
\begin{enumerate}
\item[(a)] \textbf{Affine leading order.} $y_W(\theta) = y_W(\theta)|_{\theta_j=0} + \theta_j b_j + O(\theta_j^2) + O(e^{-\lambda_f t_a})$, where $b_j$ collects the tilt, the propagated tilt forcing, and the shift of the base point.
\item[(b)] \textbf{Finite distance.} $p_j = \theta_j \|\Pi_j b_j\| + O(\theta_j^2)$, where $\Pi_j$ projects onto the orthogonal complement of the other sensitivity columns. Hence $d_j = \theta_j \|\Pi_j b_j\| + O(\theta_j^2)$, so $d_j/p_j \to 1$ as $\theta_j \to 0$, and the reduced model with $\theta_j = 0$ is a boundary point of $\mathcal{M}_W$ at distance at most $d_j$ from the current prediction.
\item[(c)] \textbf{Sloppiest direction.} $\lambda_{\min}(\mathcal{I}) \le \min_j p_j^2$. Let $u$ be a unit eigenvector of $\lambda_{\min}$ and $\lambda_2$ the next eigenvalue. Then
\[
p_j^2 = \frac{\lambda_{\min}}{u_j^2}\left(1 + O\left(\frac{\lambda_{\min}}{u_j^2 \lambda_2}\right)\right),
\]
so $p_j^2 \approx \lambda_{\min}$ if and only if, in addition to a gap $\lambda_{\min} \ll \lambda_2$, the sloppiest eigenvector is concentrated on coordinate $j$ ($u_j^2 \approx 1$). In that case the sloppiest eigenvector is, to the same order, the tangent of the compensated curve of $\theta_j$.
\item[(d)] \textbf{End point.} Taking the four null-set limits yields the zeroth-order NAIM model $\dot{x} = f(x, h_0(x))$, $y = h_0(x)$ of Proposition \ref{prop:zeroth}, with $T_{d0}' = \delta T_d$ and $T_{q0}' = \delta T_q$ if the differences are held fixed.
\end{enumerate}
\end{theorem}

\begin{proof}
(a) Lemma \ref{lem:channels} writes the output on $W$ as the graph value plus a fibre residual of order $e^{-\lambda_f t_a}$. By (B2) the graph is affine in $\theta_j$ to first order, and by (B1) the base point and the reduced flow are differentiable in $\theta_j$ at $0$. Expanding gives the result.
(b) By (a), $\partial y_W/\partial \ln \theta_j = \theta_j \partial y_W/\partial \theta_j = \theta_j b_j + O(\theta_j^2)$. Projecting orthogonally to the other columns via $\Pi_j$ gives $p_j = \theta_j \|\Pi_j b_j\| + O(\theta_j^2)$. Integrating $p_j/\theta_j'$ from $0$ to $\theta_j$ gives $d_j$, and $\Pi_j b_j \neq 0$ gives $d_j/p_j \to 1$. Since $\theta_j = 0$ lies outside the open parameter domain, the limit model is a boundary point of $\mathcal{M}_W$ at distance at most $d_j$.
(c) With eigenpairs $(\lambda_k, u^{(k)})$ of $\mathcal{I}$, $(\mathcal{I}^{-1})_{jj} = \sum_k (u_j^{(k)})^2/\lambda_k = u_j^2/\lambda_{\min} + O(1/\lambda_2)$, which yields the expansion and inequality. The tangent of the compensated curve is $\mathcal{I}^{-1}e_j = u_j u/\lambda_{\min} + O(1/\lambda_2)$.
(d) After all four limits the fibre equations are replaced by $G(x, h_0) = 0$.
\end{proof}

\subsection{Comparison with the MBAM sequence}
\begin{figure}[htbp]
\centering
\incfig{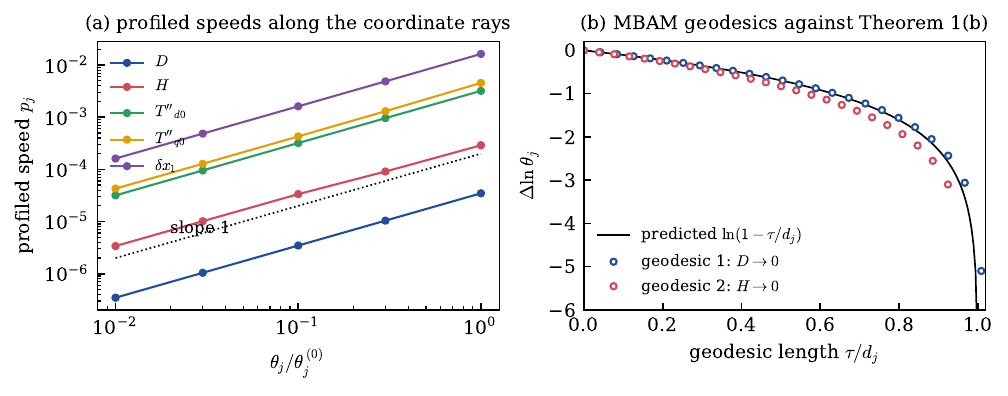}
\caption{Left: profiled speed of each MBAM parameter along its coordinate
ray, with the earlier limits imposed. Right: our two MBAM geodesics (dots)
against $\Delta\ln\theta_j = \ln(1 - \tau/d_j)$ from
Theorem~\ref{thm:boundary}(b) (lines).}
\label{fig:mbam_geod}
\end{figure}

We implemented MBAM on the same window and metric, integrating geodesics with second directional derivatives obtained by central differences (Appendix \ref{app:comp}), and evaluated the profiled speeds along each coordinate ray, removing the parameters in the published order. Table \ref{tab:mbam_comp} and Figure \ref{fig:mbam_geod} give the results.

\begin{table}[htbp]
\centering
\caption{Manifold boundaries on $[3, 5]\text{ s}$ with unit noise, in the MBAM
order of ESKT. Each row is computed with the earlier limits taken; $p_j$ and
$d_j$ as in Definition~1; slope: $d\ln p_j/d\ln\theta_j$ near the limit;
geodesic lengths and limits as in Appendix~A.}
\begin{tabular}{cccccccc}
\toprule
step & limit & free & $\sqrt{\lambda_{\min}}$ & $p_j$ & $d_j$ & slope at 0 & geodesic \\
\midrule
1 & $D \to 0$ & 11 & $3.475 \times 10^{-5}$ & $3.478 \times 10^{-5}$ & $3.47 \times 10^{-5}$ & $1.000$ & $3.52 \times 10^{-5}$ \\
2 & $H \to 0$ & 10 & $2.895 \times 10^{-4}$ & $2.900 \times 10^{-4}$ & $3.05 \times 10^{-4}$ & $0.999$ & $2.98 \times 10^{-4}$ \\
3 & $T_{d0}'' \to 0$ & 9 & $2.48 \times 10^{-3}$ & $3.20 \times 10^{-3}$ & $3.20 \times 10^{-3}$ & $1.003$ & -- \\
4 & $T_{q0}'' \to 0$ & 8 & $4.15 \times 10^{-3}$ & $4.56 \times 10^{-3}$ & $4.41 \times 10^{-3}$ & $0.996$ & -- \\
5 & $\delta x_1 \to 0$ & 7 & $1.63 \times 10^{-2}$ & $1.64 \times 10^{-2}$ & $1.63 \times 10^{-2}$ & $1.000$ & -- \\
\bottomrule
\end{tabular}
\label{tab:mbam_comp}
\end{table}

The profiled speeds are linear in $\theta_j$ to three digits as $\theta_j \to 0$ (local slope $0.996$--$1.003$). $\sqrt{\lambda_{\min}}$ also equals the profiled speed of the next boundary to within $1\%$ for $D$, $H$ and $\delta x_1$, while for $T_{d0}''$ and $T_{q0}''$ the sloppiest eigenvector mixes them ($u_{T_{d0}''}^2 = 0.49$), as Theorem \ref{thm:boundary}(c) anticipates. The ranking of the distances is what one expects MBAM to follow, since a geodesic leaves along the sloppiest eigenvector; it is not implied by the local theorem, and we checked it for the first two steps. The first two geodesics reach their boundaries at lengths $3.52 \times 10^{-5}$ and $2.98 \times 10^{-4}$, matching the local predictions $d_D = 3.47 \times 10^{-5}$ and $d_H = 3.05 \times 10^{-4}$ to within $5\%$, and the ordering $d_D < d_H < d_{T_{d0}''} < d_{T_{q0}''} < d_{\delta x_1}$ coincides with the published MBAM sequence.

\subsection{The experiment as a variable}

\begin{corollary}[The boundaries depend on the window]\label{cor:window}
For windows that start before the fibre has decayed, the $O(e^{-\lambda_f t_a})$ term in Theorem \ref{thm:boundary}(a) dominates the sensitivity to the fibre parameters, and their boundaries recede.
\end{corollary}

\begin{figure}[htbp]
\centering
\incfig{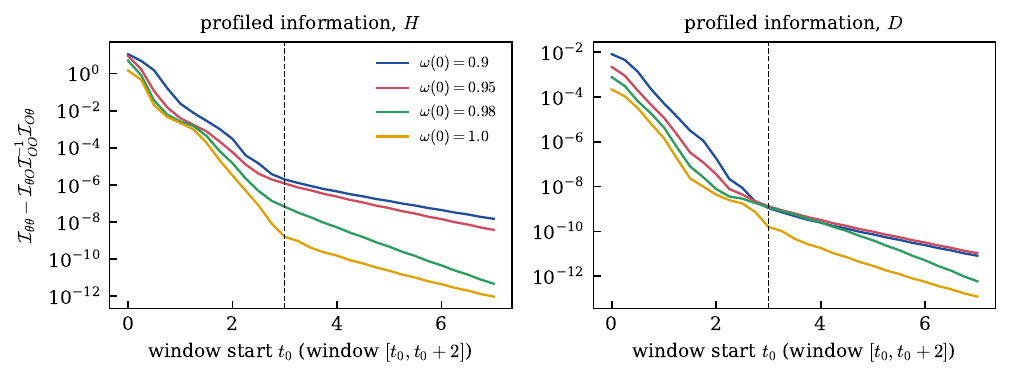}
\caption{Profiled information on $H$ (left) and $D$ (right) for windows
$[t_0, t_0 + 2]$ and four initial speeds, unit noise. Dashed: $t_0 = 3\text{ s}$
of ESKT.}
\label{fig:prof_info}
\end{figure}

Table \ref{tab:win_bounds}, Table \ref{tab:win_eig} and Figure \ref{fig:prof_info} confirm the corollary. On $[0.5, 2.5]\text{ s}$ the profiled speed of $H$ is $0.19$, about $700$ times its value on $[3, 5]\text{ s}$, and $H$ falls from second to sixth place. Starting inside the swing transient ($[0, 2]\text{ s}$) raises the count of unit-noise FIM eigenvalues above $10^{-2}$ to nine, with $H$, $T_{d0}''$ and $T_{q0}''$ among the resolved directions (Table \ref{tab:win_eig}).

\begin{table}[htbp]
\centering
\caption{Nearest boundaries for four windows, unit noise, all other
parameters free.}
\begin{tabular}{lll}
\toprule
window (s) & six smallest profiled speeds & $\sqrt{\lambda_{\min}}$ \\
\midrule
$[0.5, 2.5]$ & $D\ 8.0\text{e-}3,\ \delta x_1\ 9.2\text{e-}3,\ T_{d0}''\ 2.8\text{e-}2,\ T_{q0}''\ 5.8\text{e-}2,\ \delta T_q\ 0.15,\ H\ 0.19$ & $7.0 \times 10^{-3}$ \\
$[1, 3]$ & $D\ 2.1\text{e-}3,\ \delta x_1\ 6.7\text{e-}3,\ T_{d0}''\ 1.2\text{e-}2,\ T_{q0}''\ 1.8\text{e-}2,\ H\ 5.2\text{e-}2,\ \delta T_q\ 0.11$ & $2.1 \times 10^{-3}$ \\
$[2, 4]$ & $D\ 8.8\text{e-}5,\ H\ 3.8\text{e-}3,\ T_{q0}''\ 5.1\text{e-}3,\ T_{d0}''\ 5.2\text{e-}3,\ \delta x_1\ 5.8\text{e-}3,\ \delta T_q\ 6.8\text{e-}2$ & $8.8 \times 10^{-5}$ \\
$[3, 5]$ & $D\ 3.5\text{e-}5,\ H\ 2.6\text{e-}4,\ T_{q0}''\ 2.3\text{e-}3,\ T_{d0}''\ 2.6\text{e-}3,\ \delta x_1\ 5.2\text{e-}3,\ \delta T_q\ 1.4\text{e-}2$ & $3.5 \times 10^{-5}$ \\
\bottomrule
\end{tabular}
\label{tab:win_bounds}
\end{table}

\begin{table}[htbp]
\centering
\caption{Identifiability against window start, unit noise. Weight: squared
eigenvector mass of the parameter in the modes above $10^{-2}$.}
\begin{tabular}{lccccc}
\toprule
window (s) & $[0, 2]$ & $[0.5, 2.5]$ & $[1, 3]$ & $[2, 4]$ & $[3, 5]$ \\
\midrule
eigenvalues $> 10^{-2}$ & 9 & 7 & 7 & 6 & 5 \\
weight of $H$ in those modes & $1.00$ & $0.90$ & $0.86$ & $0.00$ & $0.00$ \\
weights of $T_{d0}'', T_{q0}''$ & $1.00, 1.00$ & $0.01, 0.10$ & $0.01, 0.14$ & $0.00, 0.00$ & $0.00, 0.00$ \\
weight of $D$ & $0.00$ & $0.00$ & $0.00$ & $0.00$ & $0.00$ \\
\bottomrule
\end{tabular}
\label{tab:win_eig}
\end{table}

Conversely, Table \ref{tab:remedies_unit} tests the two remedies conjectured in \cite{evangelou2024} on $[3, 5]\text{ s}$ under unit noise. As Lemma \ref{lem:channels} predicts, neither a larger initial speed deviation $|\omega(0) - 1|$ nor a load step that shifts the equilibrium restores $H$ or $D$: both remain at least three orders of magnitude below the cut-off $10^{-2}$.

\begin{table}[htbp]
\centering\small\setlength{\tabcolsep}{4pt}
\caption{Profiled information on $[3, 5]\text{ s}$; load steps start from
the $P_m = 0.7$ equilibrium.}
\begin{tabular}{lcccccc}
\toprule
$\omega(0)$ & $0.90$ & $0.95$ & $0.98$ (publ.) & $1.00$ & $1.02$ & $1.05$ \\
\midrule
profiled $H$ & $2.0 \times 10^{-6}$ & $1.2 \times 10^{-6}$ & $6.7 \times 10^{-8}$ & $1.6 \times 10^{-9}$ & $4.3 \times 10^{-7}$ & $7.2 \times 10^{-6}$ \\
profiled $D$ & $1.1 \times 10^{-9}$ & $1.3 \times 10^{-9}$ & $1.2 \times 10^{-9}$ & $1.6 \times 10^{-10}$ & $1.1 \times 10^{-9}$ & $9.7 \times 10^{-10}$ \\
\midrule
case & fault (publ.) & $P_m \to 0.60$ & $P_m \to 0.75$ & $P_m \to 0.80$ & $P_m \to 0.90$ & \\
\midrule
slow excursion $|\Delta e_q'|$ & $0.78$ & $0.10$ & $0.06$ & $0.14$ & $0.41$ & \\
profiled $H$ & $6.7 \times 10^{-8}$ & $1.8 \times 10^{-10}$ & $1.6 \times 10^{-10}$ & $1.3 \times 10^{-9}$ & $2.7 \times 10^{-8}$ & \\
profiled $D$ & $1.2 \times 10^{-9}$ & $2.2 \times 10^{-11}$ & $1.4 \times 10^{-11}$ & $8.0 \times 10^{-11}$ & $4.7 \times 10^{-10}$ & \\
\bottomrule
\end{tabular}
\label{tab:remedies_unit}
\end{table}

\section{Realistic measurement noise and connection to sampled ensembles}\label{sec:noise}

\subsection{Noise model and Bayesian identifiability criterion}
Table \ref{tab:inst_classes} defines three instrument classes based on IEEE Std C37.118.1 \cite{ieee_c37118,roscoe2013}. Class B (PMU steady-state accuracy limits read as $2\sigma$) is our primary physical metric.

\begin{table}[htbp]
\centering\small\setlength{\tabcolsep}{4pt}
\caption{Instrument classes (noise standard deviations).}
\begin{tabular}{p{5.3cm}ccc}
\toprule
class & $\sigma_\delta$ (rad) & $\sigma_\omega$ (pu) & $\sigma_e$ (pu), each EMF \\
\midrule
A, high grade (shaft encoder, lab acquisition) & $1 \times 10^{-3}$ & $1.0 \times 10^{-5}$ ($0.6\text{ mHz}$) & $1 \times 10^{-3}$ \\
B, PMU steady-state limits read as $2\sigma$ & $5 \times 10^{-3}$ & $4.2 \times 10^{-5}$ ($2.5\text{ mHz}$) & $5 \times 10^{-3}$ \\
C, PMU dynamic (modulation) limits read as $2\sigma$ & $1.5 \times 10^{-2}$ & $5.0 \times 10^{-4}$ ($30\text{ mHz}$) & $1.5 \times 10^{-2}$ \\
\bottomrule
\end{tabular}
\label{tab:inst_classes}
\end{table}

\begin{definition}[Physical metric and predicted recovery error]\label{def:phys_metric}
Let $\Sigma = \operatorname{diag}(\sigma^2)$ be the per-sample noise covariance of Table~\ref{tab:inst_classes}, with errors independent across states and across samples taken every $0.02\text{ s}$. The physical FIM is $\mathcal{I} = \sum_t S(t)^\top \Sigma^{-1} S(t)$. When each log-parameter is drawn uniformly from a box of width $w = \ln(1.1/0.9)$ ($\pm 10\%$ perturbation), approximating the uniform prior of the remaining parameters by a Gaussian of variance $w^2/12$ yields the van Trees (Bayesian Cram\'er--Rao) posterior standard deviation \cite{gill1995}
\[
s_j = \left[(\mathcal{I} + P_{-j})^{-1}\right]_{jj}^{1/2}, \qquad P_{-j} = \frac{12}{w^2}\operatorname{diag}(1, \dots, 0_j, \dots, 1).
\]
The predicted recovery error is the mean absolute error (MAE) of the Bayes posterior mean for a uniform variable on $[0, 1]$ observed with Gaussian error $s_j/w$. A parameter is identifiable when its predicted MAE is below $0.1$ (no information gives $0.25$). The critical noise multiple $\kappa_j^*$ is the largest factor by which all class-B standard deviations can be multiplied while the predicted MAE stays below $0.1$.
\end{definition}

\begin{proposition}[Meaning of the published cut-off]\label{prop:cutoff}
Under uniform noise $\sigma$ on every state, a FIM eigen-direction with eigenvalue $\lambda$ at $\sigma = 1$ has Cram\'er--Rao standard deviation $\sigma/\sqrt{\lambda}$ in log units. The cut-off $\lambda \ge 10^{-2}$ of \cite{evangelou2024} therefore means ``standard deviation at most $0.1$'' for $\sigma = 0.01$.
\end{proposition}

\subsection{Fisher analysis in the physical metric}
\begin{figure}[htbp]
\centering
\incfig{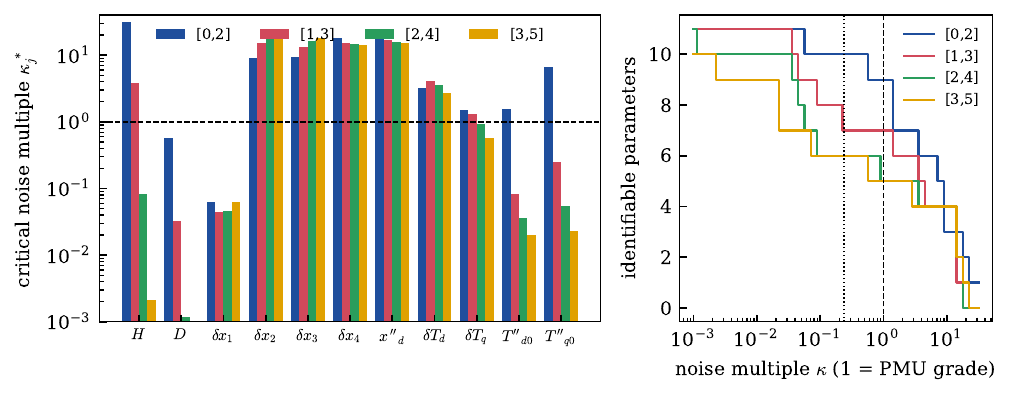}
\caption{Left: critical noise multiple $\kappa_j^*$ per parameter and window,
$\omega(0) = 0.98$; dashed: $\kappa^* = 1$. Bars at the top edge are
$\geq 31.6$; $D$ on $[3, 5]\text{ s}$ is $\leq 10^{-3}$ (search limits).
Right: identifiable parameters against the noise multiple; dotted: class A.}
\label{fig:crit_noise}
\end{figure}

Table \ref{tab:pmu_mae}, Table \ref{tab:pmu_remedies} and Figure \ref{fig:crit_noise} give the predicted recovery errors and critical noise multiples under PMU-grade noise (class B).
\begin{itemize}
\item \emph{Late window ($[3, 5]\text{ s}$).} Five parameters are identifiable at PMU-grade noise: $\delta x_2$, $\delta x_3$, $\delta x_4$, $x_d''$ and $\delta T_d$. The published sixth, $\delta T_q$, has $\kappa^* = 0.57$ (predicted MAE $0.150$) and needs instruments about twice as good. Under uniform noise $\sigma$, the published six are identifiable only for $\sigma \le 3.0 \times 10^{-3}$.
\item \emph{Early window ($[0, 2]\text{ s}$).} Nine parameters are identifiable. $H$ becomes the best-determined parameter of all ($\kappa^* \ge 31.6$, Cram\'er--Rao standard deviation $0.1\%$, MAE $0.004$). Only $\delta x_1$ and $D$ are unidentifiable at nominal conditions, with $D$ marginal (MAE $0.150$, $\kappa^* = 0.56$).
\item \emph{Conjectured remedies under realistic noise.} On $[3, 5]\text{ s}$, neither $\omega(0) = 0.90$ (MAE $0.223$ for $H$, $\kappa_H^* = 0.094$) nor load steps restore $H$ or $D$. On the early window $[0, 2]\text{ s}$, however, $\omega(0) = 0.90$ makes $D$ identifiable under class B (MAE $0.083$) and class A (MAE $0.020$). Accurate frequency measurement matters here: per-unit frequency is measured roughly a hundred times more precisely than angle, and at class B $45\%$ of the efficient information on $D$ comes from $\omega(t)$. Under class-C dynamic limits, which relax the frequency error twelvefold, $D$ is lost again ($0.217$).
\item \emph{Memory bound under noise.} For the initial-condition family on $[3, 5]\text{ s}$, the FIM with respect to $z(0)$ at class B has square-root eigenvalues $210$, $40$, $0.038$ and $0.0022$, followed by numerical round-off: a gap of three orders of magnitude after the second slow-manifold direction, confirming Lemma \ref{lem:channels}. The predicted number of recoverable initial states falls $4\to3\to2\to1$ for windows starting at $0.25$, $1$, $2$ and $3\text{ s}$.
\end{itemize}

\begin{table}[htbp]
\centering\small\setlength{\tabcolsep}{3pt}
\caption{Predicted recovery error (MAE), class B, $\omega(0) = 0.98$. Bold:
identifiable. $n_{\text{eff}} = \sum_k \lambda_k'/(1 + \lambda_k')$, with
$\lambda_k'$ the eigenvalues of $\frac{w^2}{12}\mathcal{I}$.}
\begin{tabular}{lcccccccccccc}
\toprule
window & $H$ & $D$ & $\delta x_1$ & $\delta x_2$ & $\delta x_3$ & $\delta x_4$ & $x_d''$ & $\delta T_d$ & $\delta T_q$ & $T_{d0}''$ & $T_{q0}''$ & count ($n_{\text{eff}}$) \\
\midrule
$[0, 2]$ & \textbf{.004} & .150 & .246 & \textbf{.029} & \textbf{.022} & \textbf{.009} & \textbf{.007} & \textbf{.037} & \textbf{.073} & \textbf{.070} & \textbf{.022} & 9 (9.3) \\
$[1, 3]$ & \textbf{.071} & .248 & .245 & \textbf{.021} & \textbf{.020} & \textbf{.009} & \textbf{.009} & \textbf{.031} & \textbf{.080} & .242 & .199 & 7 (7.1) \\
$[2, 4]$ & .245 & .250 & .244 & \textbf{.017} & \textbf{.017} & \textbf{.009} & \textbf{.009} & \textbf{.034} & .106 & .248 & .243 & 5 (5.8) \\
$[3, 5]$ & .250 & .250 & .244 & \textbf{.015} & \textbf{.015} & \textbf{.009} & \textbf{.008} & \textbf{.044} & .150 & .249 & .248 & 5 (5.6) \\
\bottomrule
\end{tabular}
\label{tab:pmu_mae}
\end{table}

\begin{table}[htbp]
\centering
\caption{Predicted recovery error for $H$ and $D$ under the remedies of
Table~6. Bold: identifiable.}
\begin{tabular}{lccccc}
\toprule
& \multicolumn{2}{c}{$[3, 5]\text{ s}$} & \multicolumn{2}{c}{$[0, 2]\text{ s}$} & identifiable \\
scenario (class B unless noted) & $H$ & $D$ & $H$ & $D$ & on $[3, 5]\text{ s}$ \\
\midrule
fault, $\omega(0) = 0.98$ (published) & .250 & .250 & \textbf{.004} & .150 & 5 \\
fault, $\omega(0) = 0.90$ & .223 & .250 & \textbf{.003} & \textbf{.083} & 5 \\
fault, $\omega(0) = 1.05$ & .248 & .250 & \textbf{.010} & .180 & 5 \\
fault, $\omega(0) = 0.90$, class A & .159 & .250 & \textbf{.001} & \textbf{.020} & 6 \\
fault, $\omega(0) = 0.90$, class C & .232 & .250 & \textbf{.014} & .217 & 4 \\
load step $P_m: 0.7 \to 0.6$ & .250 & .250 & \textbf{.038} & .248 & 4 \\
load step $P_m: 0.7 \to 0.8$ & .250 & .250 & \textbf{.031} & .248 & 4 \\
load step $P_m: 0.7 \to 0.9$ & .250 & .250 & \textbf{.017} & .246 & 5 \\
\bottomrule
\end{tabular}
\label{tab:pmu_remedies}
\end{table}

\subsection{Why data-driven manifold learning mirrors the Fisher spectrum}
Evangelou et al.\ \cite{evangelou2024} observed that output diffusion maps \cite{coifman2006dm} and geometric harmonics applied to $N = 10^4$ sampled trajectories identify the same parameter set as the FIM and MBAM, and noted that the agreement is not obvious: the two routes scale the data differently, and manifold learning probes finite ($\pm 10\%$) variations. Two analytical facts explain the agreement from the Fisher matrix and show when data-driven conventions distort the physical answer.

\begin{definition}[Matched-metric Fisher information]\label{def:matched_fim}
In the linear regime, rescaling each output feature $f$ to $[0, 1]$ by its empirical range $\rho_f$ across an ensemble of parameter perturbations (as done prior to Diffusion Maps in \cite{evangelou2024}) is equivalent to assuming measurement noise with standard deviation proportional to $\rho_f$. The matching Fisher information is
\[
\tilde{\mathcal{I}} = \sum_f \rho_f^{-2} s_f^\top s_f,
\]
where $s_f$ is the row of log-parameter sensitivities of feature $f$ and $\rho_f \approx 0.2 \sum_j |s_{fj}|$ for $\pm 10\%$ perturbations.
\end{definition}

Per-feature $[0, 1]$ scaling is therefore not a neutral conditioning step, but an implicit noise model. On $[3, 5]\text{ s}$, the rotor speed $\omega(t)$ varies by only about $8.5 \times 10^{-5}\text{ pu}$ ($5\text{ mHz}$). Rescaling $\omega(t)$ to $[0, 1]$ stretches this tiny tilt residual to the same unit weight as the rotor angle $\delta(t)$. In the matched metric $\tilde{\mathcal{I}}$, the relative extent $\sqrt{\tilde{\lambda}_k/\tilde{\lambda}_1}$ of the $H$-direction rises forty-fold, from $0.0004$ at $\omega(0) = 0.98$ to $0.0166$ at $\omega(0) = 0.90$, next to $\delta T_q$ ($0.0176$), and so reaches the resolution limit of a $5000$-sample ensemble. In the physical PMU metric (Table \ref{tab:pmu_remedies}) $H$ remains unidentifiable (MAE $0.223$); resolving it would require a frequency-error standard deviation of about $0.24\text{ mHz}$.

When trajectories are instead whitened by a physical noise covariance $\Sigma^{-1/2}$, the sample covariance of the trajectory ensemble is governed directly by the prior-scaled Fisher matrix:

\begin{proposition}[Spiked-covariance form of sampled trajectories]\label{prop:spiked}
Linearise the whitened output trajectory of dimension $M$ about the nominal parameter point:
\[
\zeta = \Sigma^{-1/2}(Y(\theta_0) + S\Delta) + \varepsilon,
\]
where $\Delta$ is the log-parameter perturbation with mean zero and covariance $\frac{w^2}{12}I$, and $\varepsilon \sim \mathcal{N}(0, I_M)$. The population covariance of $\zeta$ is
\[
\operatorname{Cov}(\zeta) = I_M + \frac{w^2}{12}\tilde{S}\tilde{S}^\top, \qquad \tilde{S} = \Sigma^{-1/2}S.
\]
Its eigenvalues are $1 + \lambda_k'$, where $\lambda_k'$ are the eigenvalues of the prior-scaled FIM $\frac{w^2}{12}\mathcal{I}$, and $1$ otherwise. In the Marchenko--Pastur limit $M, N \to \infty$ with $M/N \to \gamma$ \cite{marchenko1967,baik2005}, a parameter direction separates from the noise bulk $(1 + \sqrt{\gamma})^2$ if and only if $\lambda_k' > \sqrt{\gamma}$, and its sample covariance eigenvalue tends to $(1 + \lambda_k')(1 + \gamma/\lambda_k')$.
\end{proposition}

\begin{proof}
The matrices $\tilde{S}\tilde{S}^\top$ and $\tilde{S}^\top \tilde{S} = \mathcal{I}$ share their nonzero eigenvalues. The separation threshold $\lambda_k' > \sqrt{\gamma}$ and limiting eigenvalue location follow directly from the Baik--Ben Arous--P\'ech\'e phase transition for finite-rank perturbations of the identity \cite{baik2005}.
\end{proof}

In the linear regime, therefore, the leading principal directions of a whitened trajectory ensemble over a $\pm 10\%$ parameter box are the images $\tilde{S}u_k$ of the eigenvectors $u_k$ of the prior-scaled FIM $\frac{w^2}{12}\mathcal{I}$ whose eigenvalues rise above the finite-sample threshold $\sqrt{M/N}$. Once the NAIM and Fisher structure are known, a data-driven study adds a check of the linearisation rather than new information. The check succeeds for the full nonlinear model. On $[3, 5]\text{ s}$ at class B with $N = 5000$ samples, the six leading sample eigenvalues are $1.92 \times 10^4$, $1380$, $701$, $515$, $30.8$ and $2.79$, against the predicted $1.91 \times 10^4$, $1360$, $726$, $513$, $31.1$ and $2.47$; a seventh component above the edge reflects curvature of the data manifold. Geometric-harmonics regression on the components above the noise edge attains the Fisher-predicted recovery error in $113$ of $114$ parameter--scenario pairs (four windows at class B, two windows at classes A and C, $\omega(0) = 0.90$, and a family in which only $H$, $D$, $T_{d0}''$ and $T_{q0}''$ vary), with correlation $0.999$. Diffusion maps with the residual-based coordinate selection of \cite{evangelou2024} agree in $108$ of the $114$; the selection step, which can drop a weak parameter, is the fragile part.

The proposition also separates ensemble dimensionality from single-trajectory identifiability. An ensemble of $N$ simulated trajectories detects a direction whenever $\lambda_k' > \sqrt{M/N}$, a threshold that shrinks as $N$ grows, whereas a single physical record resolves a parameter only when $\lambda_k'$ is well above one. On $[3, 5]\text{ s}$ under class-B noise, $\delta T_q$ has $\lambda' = 1.25$: well above $\sqrt{M/N}$, so the ensemble sees six directions, yet the single-trajectory Bayesian MAE is $0.150$, so only five parameters are practically identifiable.

\section{Discussion and conclusions}\label{sec:discussion}

\subsection{Comparison with ESKT study}
Table \ref{tab:comparison} sets the main statements of \cite{evangelou2024} against our findings.

\begin{table}[htbp]
\centering
\small
\caption{Comparison with ESKT study.}
\begin{tabular}{>{\raggedright\arraybackslash}p{3.2cm}>{\raggedright\arraybackslash}p{5.2cm}>{\raggedright\arraybackslash}p{6.2cm}}
\toprule
item & Evangelou et al.\ \cite{evangelou2024} & this work \\
\midrule
model equations & $\omega_0 = 2\pi \cdot 60$; $i_q = (v_d - e_q'')/x_q''$ & $\omega_0 = 1\text{ pu}$; $i_q = (v_d - e_d'')/x_q''$; FIM spectrum then reproduced exactly \\
\addlinespace
identifiable set, $[3, 5]\text{ s}$ & effectively six, at cut-off $10^{-2}$ & five eigenvalues strictly above $10^{-2}$ in $\theta$ coordinates (count is coordinate-dependent); six under Bayesian criterion for uniform noise $\le 3 \times 10^{-3}$; five at PMU-grade noise \\
\addlinespace
MBAM sequence & $D, H, T_{d0}'', T_{q0}'', \delta x_1$, one geodesic at a time; first geodesic length $\approx 3.8 \times 10^{-5}$ & first four from one algebraic statement (zeroth-order null set); distances equal profiled speeds to first order; predicted vs.\ geodesic length $3.47$ vs.\ $3.52 \times 10^{-5}$; reduced model identified as $h_0$; $\delta x_1$ is a regular near-degeneracy \\
\addlinespace
time scale of $H, D$ & associated with the slowest dynamics & swing mode is a fast fibre ($\tau = 0.235\text{ s}$), $3.7\times$ faster than the faster slow flux mode \\
\addlinespace
cause of unidentifiability & $\omega(0)$ near equilibrium & fibre collapse before the window; memory confounded ($99.6$--$99.9\%$ spanned); tilt suppressed by $1/\omega_b$ \\
\addlinespace
remedy: larger $|\omega(0) - 1|$ & would identify $H, D$ & not in the late window at PMU-grade noise; identifies $D$ on $[0, 2]\text{ s}$ if steady-state frequency accuracy holds \\
\addlinespace
remedy: load change & would identify slow-variable parameters & no improvement for $H$ or $D$ \\
\addlinespace
fast sampling & needed for $T''$ & the window start is decisive, for $H$ as well; nine identifiable on $[0, 2]\text{ s}$ \\
\addlinespace
noise, scaling and data agreement & $\sigma = 1$ and $[0, 1]$ scaling as neutral conventions; FIM--data agreement non-obvious & both are implicit noise models; in the linear regime the whitened data covariance is $I_M + \frac{w^2}{12}\tilde{S}\tilde{S}^\top$ (Proposition \ref{prop:spiked}); confirmed on the nonlinear model (113/114 pairs) \\
\bottomrule
\end{tabular}

\label{tab:comparison}
\end{table}

\subsection{Conclusions}
Read through the slow-manifold geometry of the generator, the identifiability results of Evangelou et al.\ \cite{evangelou2024} have a structural explanation. The swing mode ($\tau = 0.235\text{ s}$) is $3.7$ times faster than the faster slow flux mode and has decayed by a factor $3 \times 10^{-6}$ when the $[3, 5]\text{ s}$ window opens, so $H$ and $D$ are fibre parameters rather than slow ones. The zeroth-order NAIM $h_0$ is exactly independent of $H$, $D$, $T_{d0}''$ and $T_{q0}''$, which gives four of the five MBAM limits in one statement. These parameters re-enter through the tilt in a computable order: $D$, $T_{d0}''$ and $T_{q0}''$ at first order and $H$ at second, with the swing pair further suppressed by $1/\omega_b$. After the transient, the memory of any fast parameter passes through a two-dimensional channel, the landing point on the slow manifold, which the identifiable slow parameters span to $99.94\%$ and $99.61\%$. Profiling out the slow parameters leaves only the small tilt.

Each null-set limit is a boundary of the model manifold at a distance equal, to first order in the parameter, to its current profiled speed (Theorem \ref{thm:boundary}). The ranking of these distances matches the published MBAM sequence, and the first two agree with computed geodesic lengths to within $5\%$. The terminal reduced model is the zeroth-order slow-manifold model on $h_0$, and the boundaries recede for earlier windows.

The window start and the physical noise decide the rest. At PMU-grade noise (IEEE C37.118.1), the late window $[3, 5]\text{ s}$ identifies five parameters rather than six, and neither a larger initial speed deviation nor a load change restores $H$ or $D$. The window $[0, 2]\text{ s}$ identifies nine parameters, with $H$ the best determined (Cram\'er--Rao standard deviation $0.1\%$). Rescaling trajectory features to $[0, 1]$ acts as an artificial noise model (Definition \ref{def:matched_fim}). In a consistent physical metric, the covariance of whitened trajectory ensembles is governed in the linear regime by the prior-scaled Fisher matrix (Proposition \ref{prop:spiked}), as sampled ensembles of the nonlinear model confirm.

\subsection{Limitations}
The analysis concerns one machine at one operating point, with a modest time-scale ratio (about $1/3.7$); Fenichel's hypotheses are used as a guide and their consequences checked numerically, not proved for this model. Proposition \ref{prop:zeroth} and Proposition \ref{prop:reentry}(a) are exact; the ordering of the MBAM limits beyond the first two steps was checked through profiled speeds only. The noise classes read standard limits as $2\sigma$, treat all states, including the internal EMFs, as measured, and assume errors independent across samples. Real PMU errors are correlated in time and only terminal quantities are measured, so the identifiable counts reported here are upper bounds. The Bayesian prediction uses a Gaussian approximation for the priors of the nuisance parameters and linearisation over $\pm 10\%$.

\appendix
\section{Appendix}
\subsection{Computational details}\label{app:comp}

\paragraph{Reproducibility.} Every number in Sections \ref{sec:model}--\ref{sec:noise}, apart from the slow-manifold graph computations of Section \ref{sec:late}, was produced by scripts written from the model equations alone; the scripts, their outputs and fixed random seeds are available as supplementary material.
\paragraph{Integration and sensitivities.} The model and its forward sensitivity equations in log-parameters, $\dot{S} = f_z S + f_\theta \operatorname{diag}(\theta)$ with $S = (\partial z/\partial \theta)\operatorname{diag}(\theta)$ and $S(0) = 0$, were integrated with an explicit Runge--Kutta $8(5,3)$ method (DOP853) at relative tolerance $10^{-10}$--$10^{-11}$, or, for the stiff limits of Table \ref{tab:mbam_comp}, with a BDF method using symbolically generated Jacobians; both agree with a Radau IIA reference to a relative $10^{-11}$. All Fisher quantities use sampling at $0.02\text{ s}$.

\paragraph{Slow-manifold graphs.} The iteration \eqref{eq:fraser_iter} was evaluated pointwise along the trajectory in 130-digit arithmetic. Each $h_k$ was obtained by Newton's method to a residual below $10^{-122}$. The derivatives $\partial_x h_k$ and $\partial h_k/\partial \ln \theta$ were taken by nested central differences with step $10^{-20}$.

\paragraph{Geodesics and profiled speeds.} Geodesics $\ddot{\theta} = -(J^\top J)^{-1}J^\top A_{vv}$ in the free log-parameters were integrated with an embedded Runge--Kutta $4(5)$ method (relative tolerance $10^{-5}$), starting along the sloppiest eigenvector with unit data-space speed. The second directional derivative $A_{vv}$ was taken by central differences with step $0.05$. Integration stopped when the target log-parameter had fallen by $5.1$ ($D$) or $3.1$ ($H$), with the remaining $0.6\%$ and $5.2\%$ of the length added from Theorem \ref{thm:boundary}(b). Profiled speeds were evaluated at $\theta_j/\theta_j^{(0)} \in \{1, 0.3, 0.1, 0.03, 0.01\}$ and integrated by trapezoidal quadrature. Limits already taken in Table~3 were imposed as $D=e^{-30}D^{(0)}$,
$H=e^{-5}H^{(0)}$ and $T''=e^{-7}T''^{(0)}$. 

\paragraph{Predicted recovery error.} For the rescaled error $r = s_j/w$, the Bayes posterior mean of $u \sim \mathcal{U}(0, 1)$ observed as $y = u + r\xi$ is the mean of the normal $\mathcal{N}(y, r^2)$ truncated to $[0, 1]$. Its mean absolute error was tabulated by Monte Carlo ($4 \times 10^5$ draws) on a logarithmic grid of $r \in [10^{-4}, 10^2]$ and interpolated. The critical noise multiple was found by bisection in $\log\kappa$.

\paragraph{Sampled ensembles.} Trajectories with log-parameters drawn uniformly within $\pm 10\%$ were integrated by a classical Runge--Kutta scheme with step $10^{-3}\text{ s}$ (error below $2\times10^{-9}$ against a Radau reference), sampled every $0.02\text{ s}$, given one independent noise realisation of the stated class and whitened. Principal components above the edge $(1+\sqrt{M/N})^2$, with $M = 606$, were regressed on by geometric harmonics (250 kernel eigenvectors, 80/20 train/test split, bandwidth chosen per target on an inner validation split).


\begin{thebibliography}{99}
\bibitem{evangelou2024} N. Evangelou, A. M. Stankovi\'c, I. G. Kevrekidis, and M. K. Transtrum, ``Comparing analytic and data-driven approaches to parameter identifiability: a power systems case study,'' \emph{arXiv:2412.18663}, 2024.
\bibitem{saccomani2003} M. P. Saccomani, S. Audoly, and L. D'Angi\`o, ``Parameter identifiability of nonlinear systems: the role of initial conditions,'' \emph{Automatica}, vol. 39, pp. 619--632, 2003.
\bibitem{gutenkunst2007} R. N. Gutenkunst, J. J. Waterfall, F. P. Casey, K. S. Brown, C. R. Myers, and J. P. Sethna, ``Universally sloppy parameter sensitivities in systems biology models,'' \emph{PLoS Computational Biology}, vol. 3, no. 10, e189, 2007.
\bibitem{transtrum2010} M. K. Transtrum, B. B. Machta, and J. P. Sethna, ``Why are nonlinear fits to data so challenging?'' \emph{Physical Review Letters}, vol. 104, 060201, 2010.
\bibitem{transtrum2011} M. K. Transtrum, B. B. Machta, and J. P. Sethna, ``Geometry of nonlinear least squares with applications to sloppy models and optimization,'' \emph{Physical Review E}, vol. 83, 036701, 2011.
\bibitem{transtrum2014} M. K. Transtrum and P. Qiu, ``Model reduction by manifold boundaries,'' \emph{Physical Review Letters}, vol. 113, no. 9, 098701, 2014.
\bibitem{transtrum2017} M. K. Transtrum, A. T. Sari\'c, and A. M. Stankovi\'c, ``Measurement-directed reduction of dynamic models in power systems,'' \emph{IEEE Transactions on Power Systems}, vol. 32, no. 3, pp. 2243--2253, 2017.
\bibitem{transtrum2018} M. K. Transtrum, A. T. Sari\'c, and A. M. Stankovi\'c, ``Information geometry approach to verification of dynamic models in power systems,'' \emph{IEEE Transactions on Power Systems}, vol. 33, no. 1, pp. 440--450, 2018.
\bibitem{fenichel1979} N. Fenichel, ``Geometric singular perturbation theory for ordinary differential equations,'' \emph{Journal of Differential Equations}, vol. 31, no. 1, pp. 53--98, 1979.
\bibitem{jones1995} C. K. R. T. Jones, ``Geometric singular perturbation theory,'' in \emph{Dynamical Systems}, Lecture Notes in Mathematics, vol. 1609, Springer, 1995, pp. 44--118.
\bibitem{kokotovic1986} P. V. Kokotovi\'c, H. K. Khalil, and J. O'Reilly, \emph{Singular Perturbation Methods in Control: Analysis and Design}, Academic Press, 1986.
\bibitem{singh2026} M Parimi, N. M. Singh et al., ``Parameter identifiability of grid-forming and grid-following inverters: a NAIM perspective,'' under preparation, 2026.
\bibitem{ieee_c37118} \emph{IEEE Standard for Synchrophasor Measurements for Power Systems}, IEEE Std C37.118.1-2011, 2011.
\bibitem{sauer2017} P. W. Sauer, M. A. Pai, and J. H. Chow, \emph{Power System Dynamics and Stability}, 2nd ed., Wiley-IEEE Press, 2017.
\bibitem{roussel1990} M. R. Roussel and S. J. Fraser, ``Geometry of the steady-state approximation: perturbation and accelerated convergence methods,'' \emph{Journal of Chemical Physics}, vol. 93, pp. 1072--1081, 1990.
\bibitem{kokotovic1975} P. V. Kokotovi\'c, ``A Riccati equation for block-diagonalization of ill-conditioned systems,'' \emph{IEEE Transactions on Automatic Control}, vol. 20, no. 6, pp. 812--814, 1975.
\bibitem{roscoe2013} A. J. Roscoe, ``Exploring the relative performance of frequency-tracking and fixed-filter phasor measurement unit algorithms under C37.118 test procedures,'' \emph{IEEE Transactions on Instrumentation and Measurement}, vol. 62, no. 8, pp. 2140--2153, 2013.
\bibitem{gill1995} R. D. Gill and B. Y. Levit, ``Applications of the van Trees inequality: a Bayesian Cram\'er--Rao bound,'' \emph{Bernoulli}, vol. 1, pp. 59--79, 1995.
\bibitem{coifman2006dm} R. R. Coifman and S. Lafon, ``Diffusion maps,'' \emph{Applied and Computational Harmonic Analysis}, vol. 21, pp. 5--30, 2006.
\bibitem{marchenko1967} V. A. Marchenko and L. A. Pastur, ``Distribution of eigenvalues for some sets of random matrices,'' \emph{Mathematics of the USSR-Sbornik}, vol. 1, pp. 457--483, 1967.
\bibitem{baik2005} J. Baik, G. Ben Arous, and S. P\'ech\'e, ``Phase transition of the largest eigenvalue for nonnull complex sample covariance matrices,'' \emph{Annals of Probability}, vol. 33, pp. 1643--1697, 2005.
\end{thebibliography}
\end{document}